\documentclass{emulateapj}

\usepackage{times}
\usepackage{amssymb}
\usepackage{amsmath}
\usepackage{pifont}
\usepackage{graphics}
\usepackage{xcolor}
\usepackage{epsfig}

\newcommand{\bjdtdb}{\ensuremath{\rm {BJD_{TDB}}}}
\newcommand{\feh}{\ensuremath{\left[{\rm Fe}/{\rm H}\right]}}
\newcommand{\teff}{\ensuremath{T_{\rm eff}}}
\newcommand{\logg}{${\log g}_*$}
\newcommand{\ecosw}{\ensuremath{e\cos{\omega_*}}}
\newcommand{\esinw}{\ensuremath{e\sin{\omega_*}}}
\newcommand{\msun}{\ensuremath{\,M_\Sun}}
\newcommand{\rsun}{\ensuremath{\,R_\Sun}}
\newcommand{\lsun}{\ensuremath{\,L_\Sun}}
\newcommand{\mj}{\ensuremath{\,M_{\rm J}}}
\newcommand{\rj}{\ensuremath{\,R_{\rm J}}}
\newcommand{\fave}{\langle F \rangle}
\newcommand{\fluxcgs}{10$^9$ erg s$^{-1}$ cm$^{-2}$}

\newcommand{\kms}{\,km\,s$^{-1}$}
\newcommand{\ms}{\,m\,s$^{-1}$} 

\begin{document}
\title{KELT-11\lowercase{b}: A Highly Inflated Sub-Saturn Exoplanet Transiting the $V$=8 Subgiant HD 93396}
\author{
Joshua Pepper$^{1}$, 
Joseph E. Rodriguez$^{2}$, 
Karen A. Collins$^{2,3}$,
John Asher Johnson$^{4}$, 
Benjamin J. Fulton$^{5}$, 
Andrew W. Howard$^{5}$, 
Thomas G. Beatty$^{6,7}$, 
Keivan G. Stassun$^{2,3}$,
Howard Isaacson$^{8}$, 
Knicole d. Col\'on$^{9,10}$, 
Michael B. Lund$^2$, 
Rudolf B. Kuhn$^{11}$, 
Robert J. Siverd$^{12}$, 
B. Scott Gaudi$^{13}$,
T.G. Tan$^{14}$, 
Ivan Curtis$^{15}$, 
Christopher Stockdale$^{16,17}$,
Dimitri Mawet$^{18,19}$,
Michael Bottom$^{18}$, 
David James$^{20}$, 
George Zhou$^{4}$, 
Daniel Bayliss$^{21}$, 
Phillip Cargile$^{4}$, 
Allyson Bieryla$^{4}$,
Kaloyan Penev$^{22}$,
David W. Latham$^{4}$, 
Jonathan Labadie-Bartz$^{1}$,
John Kielkopf$^{23}$,
Jason D. Eastman$^{4}$,
Thomas E. Oberst$^{24}$,
Eric L. N. Jensen$^{25}$,
Peter Nelson$^{26}$,
David H. Sliski$^{27}$,
Robert A. Wittenmyer$^{28,29,30}$,
Nate McCrady$^{31}$,
Jason T. Wright$^{6,7}$,
Howard M. Relles$^{4}$
}

\affil{$^1$Department of Physics, Lehigh University, 16 Memorial Drive East, Bethlehem, PA 18015, USA}
\affil{$^2$Department of Physics and Astronomy, Vanderbilt University, 6301 Stevenson Center, Nashville, TN 37235, USA}
\affil{$^{3}$Department of Physics, Fisk University, 1000 17th Avenue North, Nashville, TN 37208, USA}
\affil{$^4$Harvard-Smithsonian Center for Astrophysics, 60 Garden St, Cambridge, MA 02138, USA}
\affil{$^5$Institute for Astronomy, University of Hawaii, 2680 Woodlawn Drive, Honolulu, HI 96822-1839, USA}
\affil{$^{6}$Department of Astronomy \& Astrophysics, The Pennsylvania State University, 525 Davey Lab, University Park, PA 16802, USA}
\affil{$^{7}$Center for Exoplanets and Habitable Worlds, The Pennsylvania State University, 525 Davey Lab, University Park, PA 16802, USA}
\affil{$^{8}$Department of Astronomy, University of California at Berkeley, Berkeley, CA 94720, USA}
\affil{$^{9}$NASA Ames Research Center, M/S 244-30, Moffett Field, CA 94035, USA}
\affil{$^{10}$Bay Area Environmental Research Institute, 625 2nd St. Ste 209 Petaluma, CA 94952, USA}
\affil{$^{11}$South African Astronomical Observatory, Cape Town, South Africa}
\affil{$^{12}$Las Cumbres Observatory Global Telescope Network, 6740 Cortona Dr., Suite 102, Santa Barbara, CA 93117, USA}
\affil{$^{13}$Department of Astronomy, The Ohio State University, Columbus, OH 43210, USA}
\affil{$^{14}$Perth Exoplanet Survey Telescope, Perth, Australia}
\affil{$^{15}$Adelaide, Australia}
\affil{$^{16}$American Association of Variable Star Observers, 49 Bay State Rd., Cambridge, MA 02138, USA}
\affil{$^{17}$Hazelwood Observatory, Australia}
\affil{$^{18}$Department of Astronomy, California Institute of Technology, 1200 E. California Boulevard, MC 249-17, Pasadena, CA 91125, USA}
\affil{$^{19}$Jet Propulsion Laboratory, California Institute of Technology, 4800 Oak Grove Drive, Pasadena, CA 91109, USA}
\affil{$^{20}$Cerro Tololo Inter-American Observatory, La Serena, Chile}
\affil{$^{21}$Observatoire Astronomique de l'Universit\'e de Gen\`eve, 51 ch. des Maillettes, 1290 Versoix, Switzerland}
\affil{$^{22}$Department of Astrophysical Sciences, Princeton University, Princeton, NJ 08544, USA}
\affil{$^{23}$Department of Physics and Astronomy, University of Louisville, Louisville, KY 40292, USA }
\affil{$^{24}$Department of Physics, Westminster College, New Wilmington, PA 16172, USA}
\affil{$^{25}$Department of Physics and Astronomy, Swarthmore College, Swarthmore, PA 19081, USA}
\affil{$^{26}$Ellinbank Observatory, Victoria, Australia}
\affil{$^{27}$Department of Physics and Astronomy, University of Pennsylvania, Philadelphia, PA 19104, USA}
\affil{$^{28}$School of Physics, University of New South Wales, Sydney, NSW 2052, Australia}
\affil{$^{29}$Australian Center for Astrobiology, University of New South Wales, Sydney, NSW 2052, Australia}
\affil{$^{30}$Computational Engineering and Science Research Centre, University of Southern Queensland, Toowoomba, Queensland 4350, Australia}
\affil{$^{31}$Department of Physics and Astronomy, University of Montana, Missoula, MT 59812, USA}

\shorttitle{KELT-11b}

\begin{abstract}
We report the discovery of a transiting exoplanet, KELT-11b, orbiting the bright ($V=8.0$) subgiant HD 93396.  A global analysis of the system shows that the host star is an evolved subgiant star with $\teff = 5370\pm51$ K, $M_{*} = 1.438_{-0.052}^{+0.061}$\msun, $R_{*} = 2.72_{-0.17}^{+0.21}$\rsun, \logg$= 3.727_{-0.046}^{+0.040}$, and \feh$ = 0.180\pm0.075$.  The planet is a low-mass gas giant in a $P = 4.736529\pm0.00006$ day orbit, with $M_{P} =  0.195\pm0.018$\mj, $R_{P}= 1.37_{-0.12}^{+0.15}$\rj, $\rho_{P} = 0.093_{-0.024}^{+0.028}$ g cm$^{-3}$, surface gravity $\log{g_{P}} = 2.407_{-0.086}^{+0.080}$, and equilibrium temperature $T_{eq} = 1712_{-46}^{+51}$ K.  
KELT-11 is the brightest known transiting exoplanet host in the southern hemisphere by more than a magnitude, and is the 6th brightest transit host to date.   The planet is one of the most inflated planets known, with an exceptionally large atmospheric scale height (2763 km), and an associated size of the expected atmospheric transmission signal of 5.6\%.  These attributes make the KELT-11 system a valuable target for follow-up and atmospheric characterization, and it promises to become one of the benchmark systems for the study of inflated exoplanets.    

\end{abstract}


\section{Introduction}

The discovery of transiting exoplanets has been marked by two eras. The first era began with observations showing that the planet HD 209458b, first discovered by the radial velocity (RV) method, transited its host star \citep{Henry:2000, Charbonneau:2000}, and with the discovery of the first planet with the transit method, OGLE-TR-56b \citep{Udalski:2002, Konacki:2003}.  That began a period of rapid discovery of new transiting exoplanets using small, automated, and dedicated telescopes, most notably by the HATNet \citep{Bakos:2004}, SuperWASP \citep{Pollacco:2006}, TrES \citep{Alonso:2004}, and XO \citep{McCullough:2006} projects.  The second era was marked by the 2007 launch of the CoRoT mission \citep{Rouan:1998}, and then in 2009 with the launch of the Kepler mission \citep{Borucki:2010}.  Space-based detection of transiting planets was a huge leap forward, especially with the ability to detect smaller and longer-period transiting planets.

Although the Kepler mission has been tremendously fruitful with the number and variety of detected planets, especially for the determination of the underlying population and demographics of exoplanets, the small ground-based telescopes have continued to make many important discoveries.  Notably, the population of transiting planets discovered by the ground-based surveys tend to be large planets with short orbital periods orbiting bright stars, due to selection and observation bias \citep{Pepper:2003,Pepper:2005,Gaudi:2005,Pont:2006,Fressin:2007}.  These planets, unlike the vast majority of the Kepler planets, offer great potential for detailed characterization of the atmospheres of exoplanets.  The bulk of our understanding of exoplanetary atmospheres comes from observations of planets with host stars with $V < 13$ \citep{Sing:2016,Seager:2010}. 

For that reason, the ongoing discoveries from ground-based transit surveys will continue to provide great value, at least until the launch of the Transiting Exoplanet Survey Satellite (TESS) mission \citep{Ricker:2015}.  Among these projects is the KELT survey.  KELT (the Kilodegree Extremely Little Telescope) fills a niche in planet discovery space by observing stars generally brighter than those observed by the other ground-based surveys, with a target magnitude regime of $7.5 < V < 10.5$.  The KELT-North telescope \citep{Pepper:2007} has been operating since 2006, and has discovered seven exoplanets to date.  

The KELT-South telescope \citep{Pepper:2012} has been operating since 2009.  It is located in Sutherland, South Africa, and surveys a large fraction of the southern hemisphere, where no transiting planets have been discovered with a host star brighter than $V=9.2$.  KELT-South has discovered or co-discovered three transiting planets to date: KELT-10b \citep{Kuhn:2015}, and WASP-122b/KELT-14b and KELT-15b \citep{Rodriguez:2015}.  

In this paper, we report the discovery of a new exoplanet, KELT-11b.  The discovery of KELT-11b was enabled by a collaboration between the KELT team and the Retired A-star Program of the California Planet Search (CPS) team, an RV survey that has discovered 34 exoplanets \citep{Johnson:2011}.  This discovery attests to the value of combining data from multiple surveys to enable future discovery.  This planet has one of the brightest host stars in the sky for a transiting planet, and is by far the brightest transit host in the southern hemisphere.  It is also extraordinarily inflated, and one of the lowest-density planets known.

\section{Discovery and Follow-Up Observations}

In this section, we describe the KELT data and follow-up photometric and RV observations that led to the discovery and confirmation of KELT-11b. A detailed description of the KELT pipeline and candidate selection process can be found in \citet{Siverd:2012} and \citet{Kuhn:2015}.

\subsection{KELT-South} \label{sec:ks}
\begin{figure*}[ht]
\centering\epsfig{file=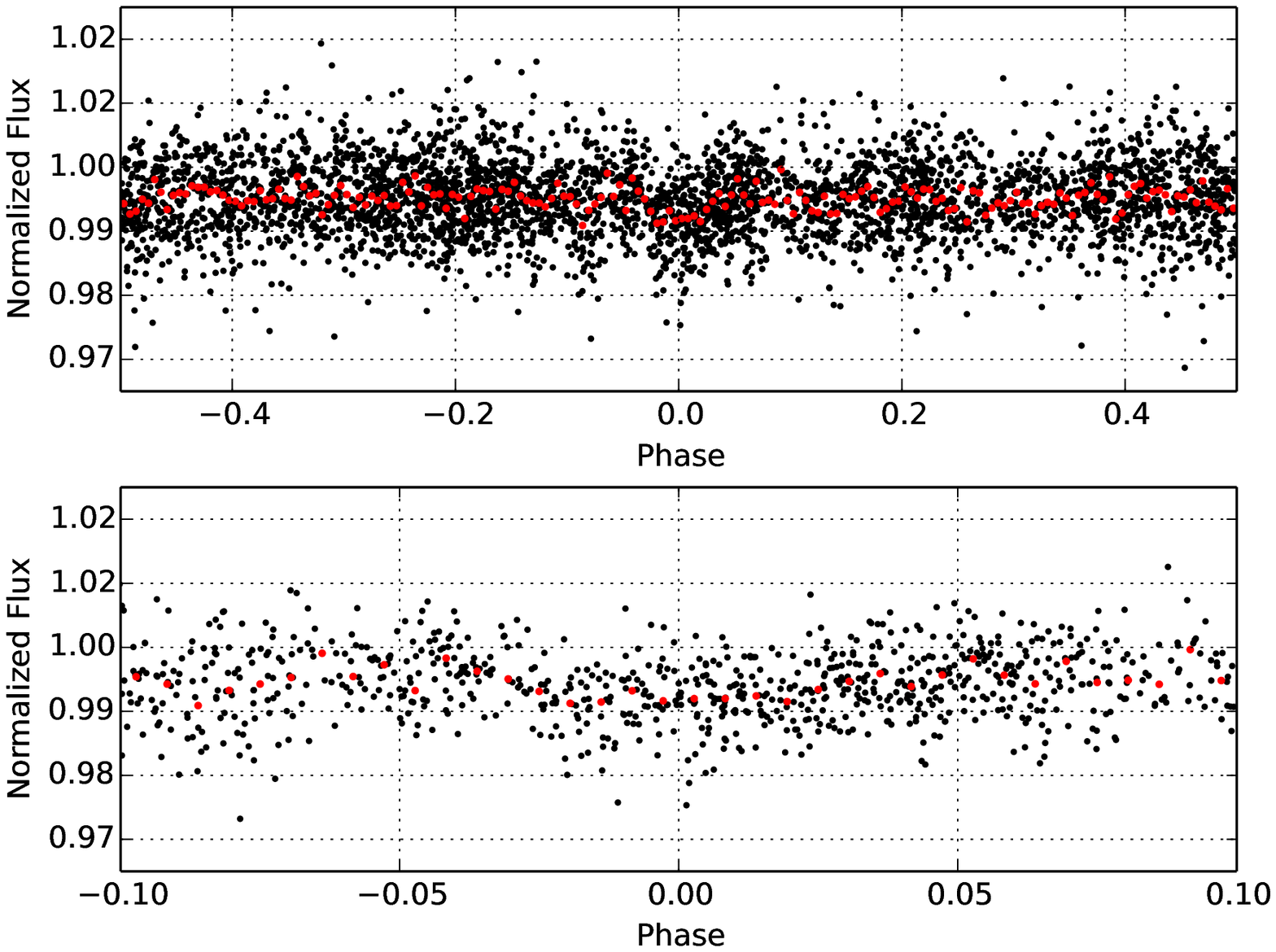,clip=,width=0.99\linewidth}
\caption{Discovery light curve of KELT-11b from the KELT-South telescope. The light curve contains 3910 observations spanning just over 4 years, phase folded to the orbital period of P = 4.7360001 days; the red points show the light curve binned in phase using a bin size of 0.005 in phase.}
\label{fig:DiscoveryLC}
\end{figure*}

KELT-11b is located in the KELT-South field 23, which is centered at J2000 $\alpha =$ 10$^{h}$ 43$^{m}$ 48$^{s}$, $\delta =$ -20$\degr$ 00$\arcmin$ 00$\arcsec$. This field was monitored from UT 2010 March 12 to UT 2014 July 9, resulting in 3910 images after post-processing and removal of bad images. Following the same strategy as described in \citet{Kuhn:2015}, we reduced the raw images, extracted the light curves, and searched for transit candidates. One candidate from this processing, KS23C00790 (HD 93396, TYC 5499-1085-1, 2MASS J10464974-0923563), located at $\alpha =$ 10$^{h}$ 46$^{m}$ 49$\fs$7, $\delta =$ -9$\degr$ 23$\arcmin$ 56$\arcsec$5 J2000, was rated as a top candidate from the field. The host star properties can be seen in Table \ref{tab:Host_Lit_Props}. We use the Box-fitting Least-squares algorithm \citep[BLS;][]{Kovacs:2002} as implemented in the \textsc{VARTOOLS} software package \citep{Hartman:2016} to identify transit candidates, making use of statistics that come from the \textsc{VARTOOLS} package, as well as from \citet{Pont:2006} and \citet{Burke:2006}.  The associated selection criteria and the corresponding values for KELT-11b are shown in Table \ref{tab:Selection_Criteria}.  The discovery light curve itself is shown in Figure \ref{fig:DiscoveryLC}.

\begin{table*}
\centering
\caption{Stellar Properties of KELT-11}
\label{tab:Host_Lit_Props}
\begin{tabular}{llccc}
   \hline
  \hline
\hline
  Parameter & Description & KELT-11 Value & Source & Reference(s) \\
 Names 			& 					& HD 93396 		& 		&			\\
			& 					&  TYC 5499-1085-1	& 		&			\\
			& 					& 2MASS J10464974-0923563 		& 		&			\\
			&					&				&		&			\\
$\alpha_{J2000}$	&Right Ascension (RA)& 10:46:49.74065			& Tycho-2	& \citet{Hog:2000}	\\
$\delta_{J2000}$	&Declination (Dec)& -09:23:56.4870			& Tycho-2	& \citet{Hog:2000}	\\
$B_T$			&Tycho B$_T$ magnitude& 9.072 $\pm$ 0.018		& Tycho-2	& \citet{Hog:2000}	\\
$V_T$			&Tycho V$_T$ magnitude& 8.130 $\pm$ 0.013		& Tycho-2	& \citet{Hog:2000}	\\
$V$			&Johnson V magnitude& 8.03 		& Hipparcos	& \citet{ESA:1997}	\\
			&					&				&		&			\\
$J$			&2MASS magnitude& 6.616 $\pm$ 0.024		& 2MASS 	& \citet{Cutri:2003, Skrutskie:2006}	\\
$H$			&2MASS magnitude& 6.251 $\pm$ 0.042		& 2MASS 	& \citet{Cutri:2003, Skrutskie:2006}	\\
$K_S$			&2MASS magnitude& 6.122 $\pm$ 0.018		& 2MASS 	& \citet{Cutri:2003, Skrutskie:2006}	\\
			&					&				&		&			\\
\textit{W1}		&WISE passband& 6.152 $\pm$ 0.1		& WISE 		&\citet{Wright:2010, Cutri:2014}	\\
\textit{W2}		&WISE passband& 6.068 $\pm$ 0.036		& WISE 		& \citet{Wright:2010, Cutri:2014}\\
\textit{W3}		&WISE passband& 6.157 $\pm$ 0.015		& WISE 		& \citet{Wright:2010, Cutri:2014}	\\
\textit{W4}		&WISE passband& 6.088	$\pm$ 0.048			& WISE 		& \citet{Wright:2010, Cutri:2014}	\\
			&					&				&		&			\\
$\mu_{\alpha}$		& Proper Motion in RA (mas yr$^{-1}$)	& -78.30 $\pm$ 0.95 		& Tycho-2		& \citet{Leeuwen:2007} \\
$\mu_{\delta}$		& Proper Motion in DEC (mas yr$^{-1}$)	& -77.61 $\pm$ 0.68		& Tycho-2		& \citet{Leeuwen:2007} \\
			&					&				&		&			\\
U$^{*}$ & Space motion (\kms) &  -9.6 $\pm$ 1.2  & &  This work \\
V & Space motion (\kms) & -46.2 $\pm$ 3.0 & & This work \\
W & Space motion (\kms) & -8.4 $\pm$ 3.4  &  &  This work \\
Distance & Distance (pc) & 98 $\pm$ 5 &  &  This work \\
RV & Absolute RV (\kms) &  35.0 $\pm$ 0.1  & &  This work \\
$v\sin{i_*}$&  Stellar Rotational Velocity (\kms)    &  $2.66 \pm 0.50$ & & This work \\
 \hline
\hline
\hline
\end{tabular}

 \footnotesize \textbf{\textsc{NOTES}}\\

\footnotesize $^{*}$U is positive in the direction of the Galactic Center 
\end{table*}

\begin{table}
 \caption{KELT-South selection criteria for candidate KS23C00790.}
\small
\label{tab:Selection_Criteria}
 \begin{tabular}{lll}
    \hline
    Statistic &  Selection & Value for \\
    & Criteria& KS23C00790 / KELT-11\\
    \hline
    Signal detection efficiency\dotfill & SDE $>$ 7.0 & 9.91663 \\
    Signal to pink-noise\dotfill & SPN $>$ 7.0 & 8.16926\\
    Transit depth\dotfill & $\delta <$ 0.05 & 0.00344\\
    $\chi^2$ ratio\dotfill & $\displaystyle\frac{\Delta\chi^2}{\Delta\chi^2_{-}} >$ 1.5 & 1.808\\
    Duty cycle\dotfill & q $<$ 0.1 & 0.05333\\
    \hline
    
 \end{tabular}
\end{table}  
\vspace{1in}

\subsection{Photometric Follow-up}
\label{sec:Follow-up_Photometry}
We obtained follow-up time-series photometry of KELT-11b to check for false positives such as eclipsing binary stars, and to better determine the transit shape, depth, and duration. We used the TAPIR software package \citep{Jensen:2013} to predict transit events, and we obtained 9 full or partial transits in multiple bands between 2015 January and 2016 February. All data were calibrated and processed using the AstroImageJ package (AIJ)\footnote{http://www.astro.louisville.edu/software/astroimagej} \citep{Collins:2013,Collins:2016} unless otherwise stated.  The follow-up light curves are displayed in Figure \ref{fig:All_Lightcurve}, and Table \ref{tab:followup_lcs} lists the details of the observations.  A binned combination of all of the follow-up light curves is shown in the bottom panel of Figure \ref{fig:All_Lightcurve}, although that combination is for display purposes, and not used for analysis.

It was extremely difficult to obtain reliable ground-based photometry, due to the combination of the shallow transit depth (about 2.5 mmag) and long duration (7.3 hours).  Any given observatory will have difficulty observing a full transit including out-of-transit baseline observations within a single night.  In fact, KELT-11b is the longest-duration transiting planet discovered using the transit method from a ground-based facility, and the one with the shallowest transit as well.  We managed to obtain one full transit, and a number of partial transits that covered ingresses and egresses.

\subsubsection{Westminster College Observatory}

 We observed an ingress of KELT-11b from the Westminster College Observatory (WCO), PA, on UT 2015 January 1 in the $I$ filter. The observations employed a 0.35 m f/11 Celestron C14 Schmidt-Cassegrain telescope and SBIG STL-6303E CCD with a $\sim$ 3k $\times$ 2k array of 9 $\mu$m pixels, yielding a $24\arcmin \times 16\arcmin$ field of view and 1.4$\arcsec $pixel$^{-1}$ image scale at $3 \times 3$ pixel binning.

\subsubsection{MORC}

We observed a partial transit of KELT-11b using an 0.6m RCOS telescope at Moore Observatory (MORC), operated by the University of Louisville. The telescope has an Apogee U16M 4K $\times$ 4K CCD, giving a 26$\arcmin$ $\times$ 26$\arcmin$ field of view and 0.39$\arcsec$ pixel$^{-1}$.  We observed the transit on UT 2015 February 08 in alternating Sloan $g$ and $i$ filters from before the ingress and past the mid-transit.

\subsubsection{MINERVA}

We observed a transit of KELT-11b in the Sloan $i-$band using one of the MINERVA Project telescopes \citep{Swift:2015} on the night of UT 2015 February 08. MINERVA uses four 0.7m PlaneWave CDK-700 telescopes that are located on Mt. Hopkins, AZ, at the Fred L. Whipple Observatory. While the four telescopes are normally used to feed a single spectrograph to discover and characterize exoplanets through RV measurements, for the KELT-11 observations, we used a single MINERVA telescope in its photometric imaging mode.  That telescope had an Andor iKON-L 2048$\times$2048 camera, which gave a field of view of 20.9$\arcmin$ $\times$ 20.9$\arcmin$ and a plate scale of $0.6\arcsec$ pixel$^{-1}$. The camera has a $2048$ $\times$ $2048$ back-illuminated deep depletion sensor with fringe suppression. Due to the brightness of KELT-11 we heavily defocused for our observations, such that the image of KELT-11 was a ``donut'' approximately 20 pixels in diameter. 

\subsubsection{PEST Observatory}

On UT 2015 March 08, we observed a partial transit from the Perth Exoplanet Survey Telescope (PEST) Observatory, located in Perth, Australia.  The observations were taken with a 0.3m Meade LX200 telescope working at f/5, and with a 31' $\times$ 21' field of view.  The camera is an SBIG ST-8XME, with $1530 \times 1020$ pixels, yielding 1.2$\arcsec$ pixel$^{-1}$.  An ingress was observed using a Cousins $I$ filter. 

\subsubsection{Ivan Curtis Observatory}

On UT 2015 March 03, we observed a partial transit at the Ivan Curtis Observatory (ICO), located in Adelaide, Australia.  The observations were taken with a 0.235m Celestron Schmidt-Cassegrain telescope with an Antares 0.63x focal reducer, giving an overall focal ratio of f/6.3.  The camera is an Atik 320e, which uses a cooled Sony ICX274 CCD of $1620 \times 1220$ pixels. The field of view is 16.6' $\times$ 12.3', with a resolution of 0.62$\arcsec$ pixel$^{-1}$.  An egress was observed using a Johnson $R$ filter.

\subsubsection{Peter van de Kamp Observatory}

We observed an ingress in the Sloan $z$-band at the Swarthmore College Peter van de Kamp Observatory (PvdK) on 2015 March 18. The observatory uses a 0.6m RCOS Telescope with an Apogee U16M 4K $\times$ 4K CCD, giving a $26\arcmin \times 26\arcmin$ field of view. Using $2 \times 2$ binning, it has 0.76$\arcsec$ pixel$^{-1}$.

\subsubsection{LCOGT-CPT}

We observed an egress of KELT-11b in the Sloan $i$-band during bright time on UT 2015 May 04, using one of the 1-m telescopes in the Las Cumbres Observatory Global Telescope (LCOGT) network$\footnote{http://lcogt.net/}$ located at the South African Astronomical Observatory (SAAO) in Sutherland, South Africa.  The LCOGT telescopes at SAAO have 4K $\times$ 4K SBIG Science cameras and offer a 16$\arcmin$ $\times$ 16$\arcmin$ field of view and an unbinned pixel scale of 0.23$\arcsec$ pixel$^{-1}$.

%

\subsubsection{MVRC}

We observed one full transit of KELT-11b using the Manner-Vanderbilt Ritchey-Chr\'{e}tien (MVRC) telescope located at the Mt. Lemmon summit of Steward Observatory, AZ, on UT 2016 February 22 in the $r'$ filter. The observations employed a 0.6 m f/8 RC Optical Systems Ritchey-Chr\'{e}tien telescope and SBIG STX-16803 CCD with a $\sim$ 4k $\times$ 4k array of 9 $\mu$m pixels, yielding a $26\arcmin \times 26\arcmin$ field of view and 0.39$\arcsec $pixel$^{-1}$ image scale. The telescope was heavily defocused, resulting in a typical ``donut'' shaped stellar PSF with a diameter of $\sim$ 25$\arcsec$.

\begin{table*}
 \centering
 \caption{Photometric follow-up observations and the detrending parameters used by AIJ for the global fit.}
 \label{tab:followup_lcs}
 \begin{tabular}{llllllll}
    \hline
    \hline
    Observatory & Date (UT) & Filter & FOV & Pixel Scale & Exposure (s)  & Detrending parameters for global fit \\
    \hline
    WCO         & UT 2015 January 01    & $I$           & 24$\arcmin \times 16\arcmin$          & 1.4$\arcsec$  & binned 4 $\times$ 10 & Airmass, Time \\
    MORC        & UT 2015 February 08   & $g^{\prime}$  & 26.6$\arcmin$ $\times$ 26.6$\arcmin$  & 0.39$\arcsec$ & 20 & Airmass, Time \\
    MORC        & UT 2015 February 08   & $i^{\prime}$  & 26.6$\arcmin$ $\times$ 26.6$\arcmin$  & 0.39$\arcsec$ & 20 & Airmass, Time \\
    MINERVA     & UT 2015 February 08   & $i^{\prime}$  & 20.9$\arcmin$ $\times$ 20.9$\arcmin$  & 0$\arcsec.6$  &  6 & Airmass, Time\\
    PEST        & UT 2015 March 08      & $I$           & 31$\arcmin$ $\times$ 21$\arcmin$      & 1.2$\arcsec$  & binned 8 $\times$ 15 & Airmass, Time \\
    ICO         & UT 2015 March 13      & $R$           & 16.6$\arcmin$ $\times$ 12.3$\arcmin$  & 0.62$\arcsec$ & 20 & Airmass, Time\\
    PvdK        & UT 2015 March 18      & $z^{\prime}$  & 26$\arcmin$ $\times$ 26$\arcmin$      & 0.76$\arcsec$ & 45 & Airmass, Time\\
    LCOGT (CPT) & UT 2015 May 04        & $i^{\prime}$  &15.8$\arcmin$ $\times$ 15.8$\arcmin$   & 0.23$\arcsec$ & 12 & Airmass, Time\\
    MVRC        & UT 2016 February 22   & $r^{\prime}$  & 26$\arcmin$ $\times$ 26$\arcmin$      & 0.39$\arcsec$  & binned 6 $\times$ 30 & Airmass\\
     \hline
    \hline
 \end{tabular}
\begin{flushleft}
  \footnotesize \textbf{\textsc{NOTES}} \\
  \footnotesize All the follow-up photometry presented in this paper is available in machine-readable form in the online journal.
  \end{flushleft}
\end{table*}  

\begin{figure}
  \centering
\includegraphics[width=1\linewidth,clip,height=5.2in]{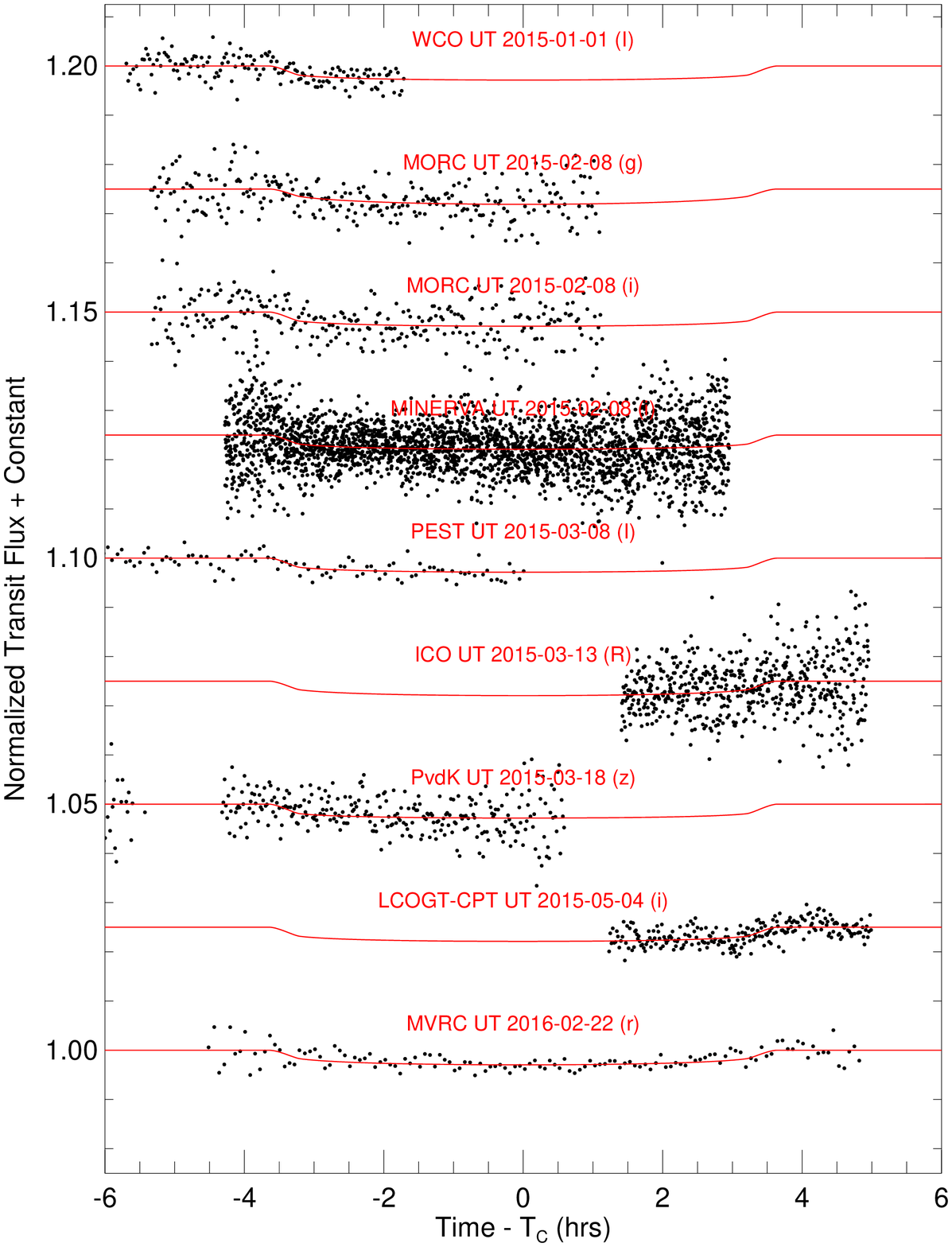}

\vspace{-.1in}
\centering\includegraphics[width=1\linewidth,clip]{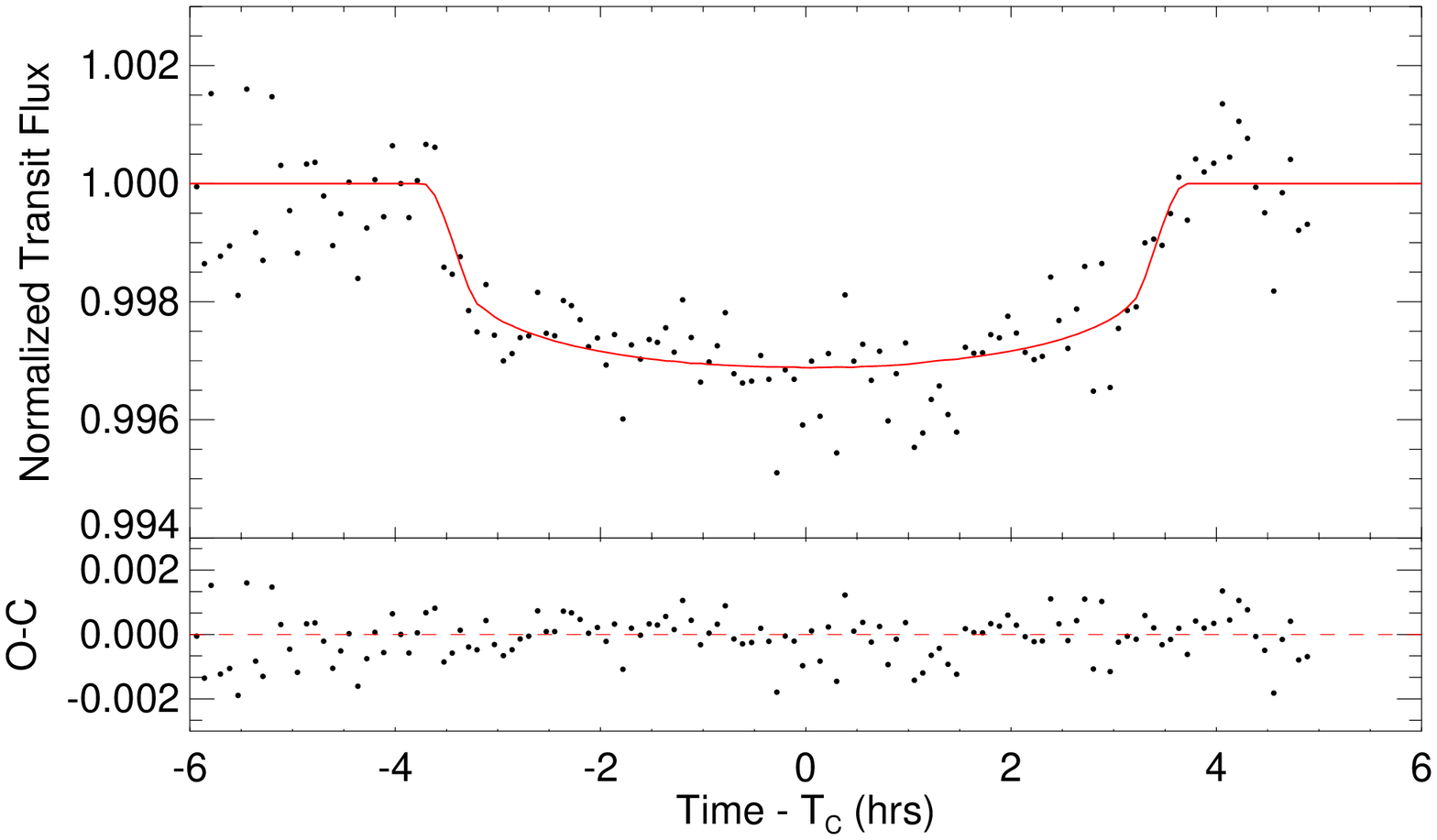}
\caption{(Top) The individual follow-up observations of KELT-11b from the KELT Follow-up Network.  The global model with limb darkening coefficients corresponding to the filter band is over-plotted on each light curve in red.  (Bottom) All follow-up transit combined and binned in 5 minute intervals to best represent the transit features. This plot is not used in the analysis and is only for display. The combined and binned models from each transit are represented by the red line.}
\label{fig:All_Lightcurve} 
\end{figure}


\subsection{Spectroscopic Follow-up}\label{sec:rvs}

\begin{figure*}
\includegraphics[width=1\linewidth]{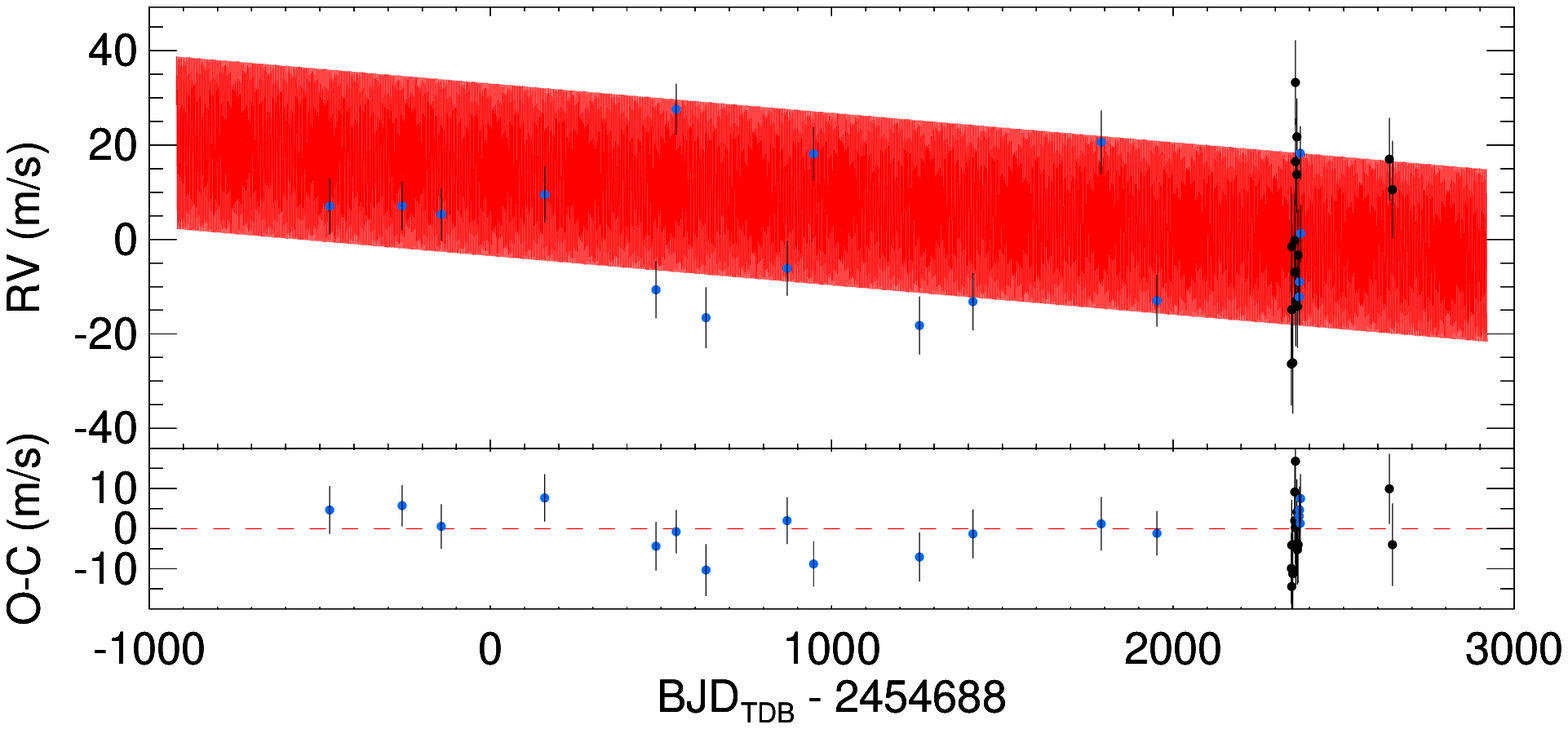}
   \vspace{-.3in}
\includegraphics[width=1\linewidth]{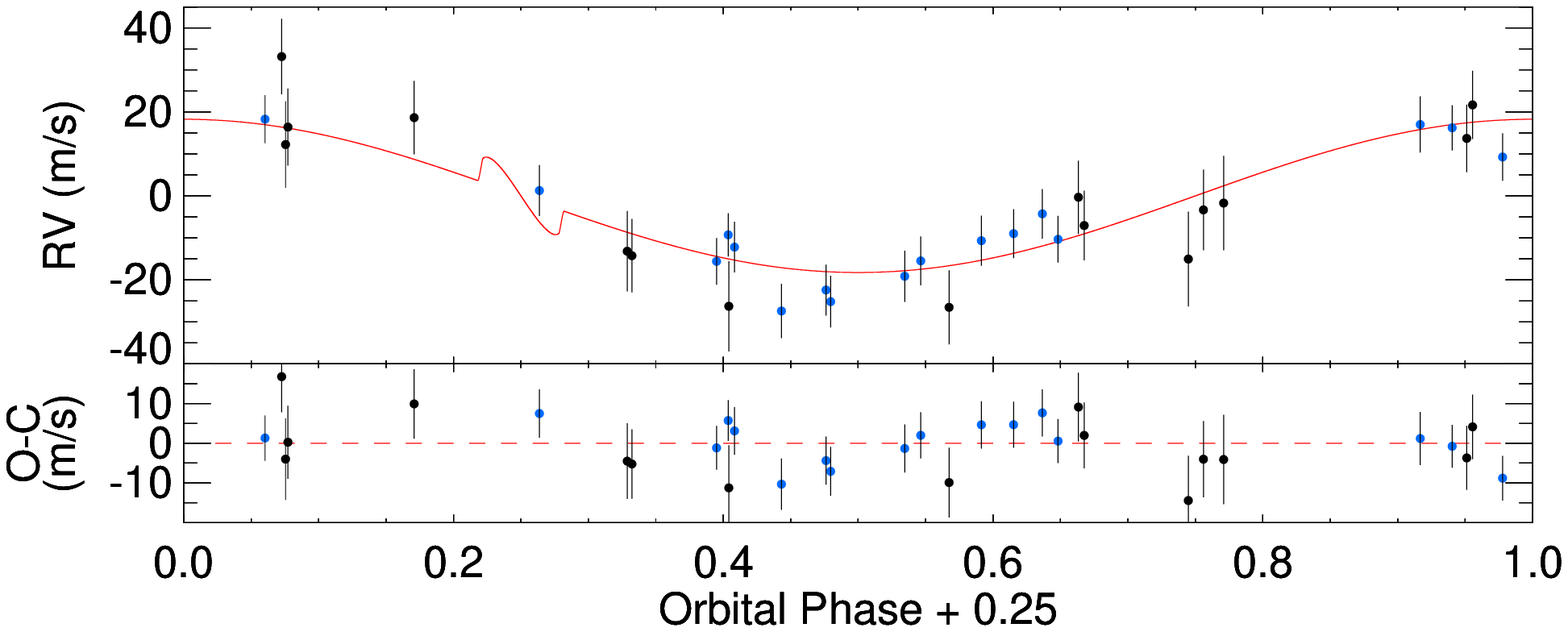}
   \vspace{.1in}
\caption{(Top) The RV measurements of KELT-11 by HIRES (blue) and APF (black). The absolute median RV has been subtracted off. The residuals of the best fit orbit (red) are shown below. (Bottom) The RV measurements of KELT-11 phase-folded to the period from our global model, adopting the circular fit (See \S \ref{sec:exofast}){, with the absolute median RV and slope subtracted}.  The predicted Rossiter-McLaughlin effect assumes perfect spin-orbit alignment and it is not constrained by our data.}
\label{fig:RVs} 
\end{figure*}

We obtained spectroscopic observations of KELT-11 to measure the RV orbit of the planet, and to measure the parameters of the host star.  The observations that were used to derive stellar parameters are listed in Table \ref{tab:spectroscopic_parameters}.  The observations that provide RV measurements are listed in Table \ref{tab:RVs}, and are displayed in Figure \ref{fig:RVs}.

\begin{table*}
 \centering
 \caption{Spectroscopic follow-up observations}
 \label{tab:spectroscopic_parameters}
 \begin{tabular}{lllllll}
    \hline
    \hline
Telescope/Instrument & Date Range & Resolution & Wavelength Range & Mean SNR & Epochs\\
    \hline
FLWO 1.5m/TRES & UT 2015 January 28 & $\approx$44,000 & 3900 - 8900\AA& $\sim$100 & 1\\
KECK/HIRES & UT 2007 April 26 -- UT 2015 February 08 &  $\approx$55,000 & 3640 - 7990\AA &$\sim$150 & 16\\
APF/Levy & UT 2015 January 12 -- UT 2015 November 4 &  $\approx$100,000 & 3740 - 9700\AA&$\sim$100  & 16 \\
   \hline
    \hline
 \end{tabular}
\end{table*}  


\subsubsection{TRES} \label{sec:tres}

In order to measure the host star properties, we obtained a spectrum with the Tillinghast Reflector Echelle Spectrograph (TRES), on the 1.5 m telescope at the Fred Lawrence Whipple Observatory (FLWO) on Mt. Hopkins, Arizona, on UT 2015 January 28. The spectrum has a resolution of $R = 44,000$, signal-to-Noise ratio (SNR) $=100.4$, and was extracted as described in \citet{Buchhave:2010}.

\begin{table}
 \centering
 \caption{Stellar parameters from the spectral analysis}
 \label{tab:stellar_params}
 \begin{tabular}{lcccc}
    \hline
    \hline
Parameter  & TRES & KECK HIRES & APF\\
    \hline
$\teff$ (K) &  $5390 \pm 50$  &$5413\pm60$ & $5444\pm60$  \\
\logg & $3.79 \pm 0.10$ &$3.87\pm0.08$ &$3.86\pm0.08$ \\
\feh &  $0.19 \pm 0.08^a$ &$0.31\pm0.04$& $0.37\pm0.04$  \\
   \hline
    \hline
 \end{tabular}
\begin{flushleft}
\footnotesize $^a$ The SME analysis used for the TRES spectra yields a value for bulk metallicity [m/H] rather than \feh.
\end{flushleft}
\end{table}  

\subsubsection{KECK HIRES} \label{sec:keck}

Well before KELT observations of this star began, the RV of HD 93396 had been monitored at Keck Observatory using the High Resolution Echelle Spectrometer \citep[HIRES,][]{Vogt:1994} starting in 2007 as part of the ``Retired A Stars" program \citep{Johnson:2006,Johnson:2011}.  Observations were conducted using the standard setup of the California Planet Survey \citep{Howard:2010,Johnson:2010} using the B5 decker and the iodine cell. RV measurements were made with respect to a high signal-to-noise ratio, iodine-free template observation \citep{Butler:1996}, which we also use to measure the stellar properties. Exposure times ranged from 50 seconds to 120 seconds depending on the seeing, with an exposure meter ensuring that all exposures reached SNR~$ \approx 150$ per pixel at 550~nm. 

\subsubsection{APF} \label{sec:apf}

To supplement the HIRES RV spectra, we also observed KELT-11 with the Levy spectrograph on the Automated Planet Finder (APF) telescope at Lick Observatory.  We collected 16 RV measurements between 2015 January 12 and 2015 November 4. The observational setup was similar to the setup used for the APF observations described in \citet{Fulton:2015}. We observed the star through a cell of gaseous iodine using the standard 1x3$\arcsec$ slit for a spectral resolution of R$\approx$100,000, and collected an iodine-free template spectrum using the 0.75$\arcsec$ $\times$ 8$\arcsec$ slit \citep[R$\approx$120,000,][]{Vogt:2014}. As with the Keck/HIRES velocities, the RVs are measured using a forward modeling process that fits for the Doppler shift of the deconvolved stellar template with respect to the stationary forest of molecular iodine absorption lines while simultaneously modeling the instrumental point spread function \citep{Butler:1996}. The photon-weighted mid-exposure times are calculated using a software-based exposure meter that measures the real-time photon flux received on the guider camera during each spectral exposure \citep{Kibrick:2006}. Exposure times ranged from 18 to 30 minutes depending on seeing and transparency to obtain SNR$\approx$100 pixel$^{-1}$ at 550 nm.

\begin{table}
 \centering
 \setlength\tabcolsep{1.5pt}
\caption{KELT-11 RV observations with HIRES and APF.}
 \begin{tabular}{cccccc}
 \hline
 \hline
  $\bjdtdb$ & RV & RV error & Bisector & Bisector Error & Instrument\\
   & (\ms)&  (\ms) & (\ms) & (\ms) & \\
 \hline
2454216.875149 & 4.4 & 1.5 & -- & -- & HIRES\\
2454429.129861 & 4.5 & 1.3 & -- & -- & HIRES\\
2454543.964648 & 2.7 & 1.4 & -- & -- & HIRES\\
2454847.046979 & 6.9 & 1.5 & -- & -- & HIRES\\
2455173.107559 & -13.3 & 1.5 & -- & -- & HIRES\\
2455232.144465 & 25.0 & 1.4 & -- & -- & HIRES\\
2455319.783317 & -19.2 & 1.6 & -- & -- & HIRES\\
2455557.098633 & -8.7 & 1.5 & -- & -- & HIRES\\
2455634.925941 & 15.5 & 1.4 & -- & -- & HIRES\\
2455945.177177 & -20.9 & 1.6 & -- & -- & HIRES\\
2456101.742516 & -15.8 & 1.5 & -- & -- & HIRES\\
2456477.738143 & 18.0 & 1.7 & -- & -- & HIRES\\
2456641.045559 & -15.6 & 1.4 & -- & -- & HIRES\\
2457057.922673 & -14.8 & 1.5 & -- & -- & HIRES\\
2457058.902596 & -11.6 & 1.5 & -- & -- & HIRES\\
2457061.010742 & 15.6 & 1.4 & -- & -- & HIRES\\
2457061.974167 & -1.4 & 1.5 & -- & -- & HIRES\\
\hline
2457034.993517 & -24.0 &  2.9 &   2.5 &  2.6 & APF\\
2457035.833572 & -12.5 &  3.7 &  -4.8 &  6.7 & APF\\
2457035.957552 &   0.8 &  3.7 &  -9.6 &  6.8 & APF\\
2457038.956627 & -23.8 &  3.5 &  -4.8 &  4.8 & APF\\
2457044.920751 &   2.2 &  2.9 &   2.5 &  2.2 & APF\\
2457044.941030 &  -4.6 &  2.7 &   3.1 &  4.7 & APF\\
2457046.859494 &  35.6 &  2.9 &  13.4 &  8.8 & APF\\
2457046.881613 &  18.9 &  3.0 &  -1.3 &  7.3 & APF\\
2457048.072907 & -10.7 &  3.1 &  -5.8 &  5.2 & APF\\
2457051.020767 &  16.1 &  2.6 &   5.8 &  4.7 & APF\\
2457051.041243 &  24.1 &  2.7 &  -1.1 &  2.8 & APF\\
2457052.825840 & -11.8 &  2.9 &   7.1 &  7.1 & APF\\
2457054.832772 &  -0.9 &  3.2 &   4.7 &  4.3 & APF\\
2457322.043143 &  19.4 &  2.9 &   4.8 &  1.7 & APF\\
2457331.064594 &  12.9 &  3.4 & -20.7 &  6.7 & APF\\
\hline
 \hline
\end{tabular}
 \label{tab:RVs}
\begin{flushleft}
\end{flushleft}
\end{table}

\subsubsection{Bisector Spans} \label{sec:bis}

\begin{figure}
\includegraphics[width=1\linewidth]{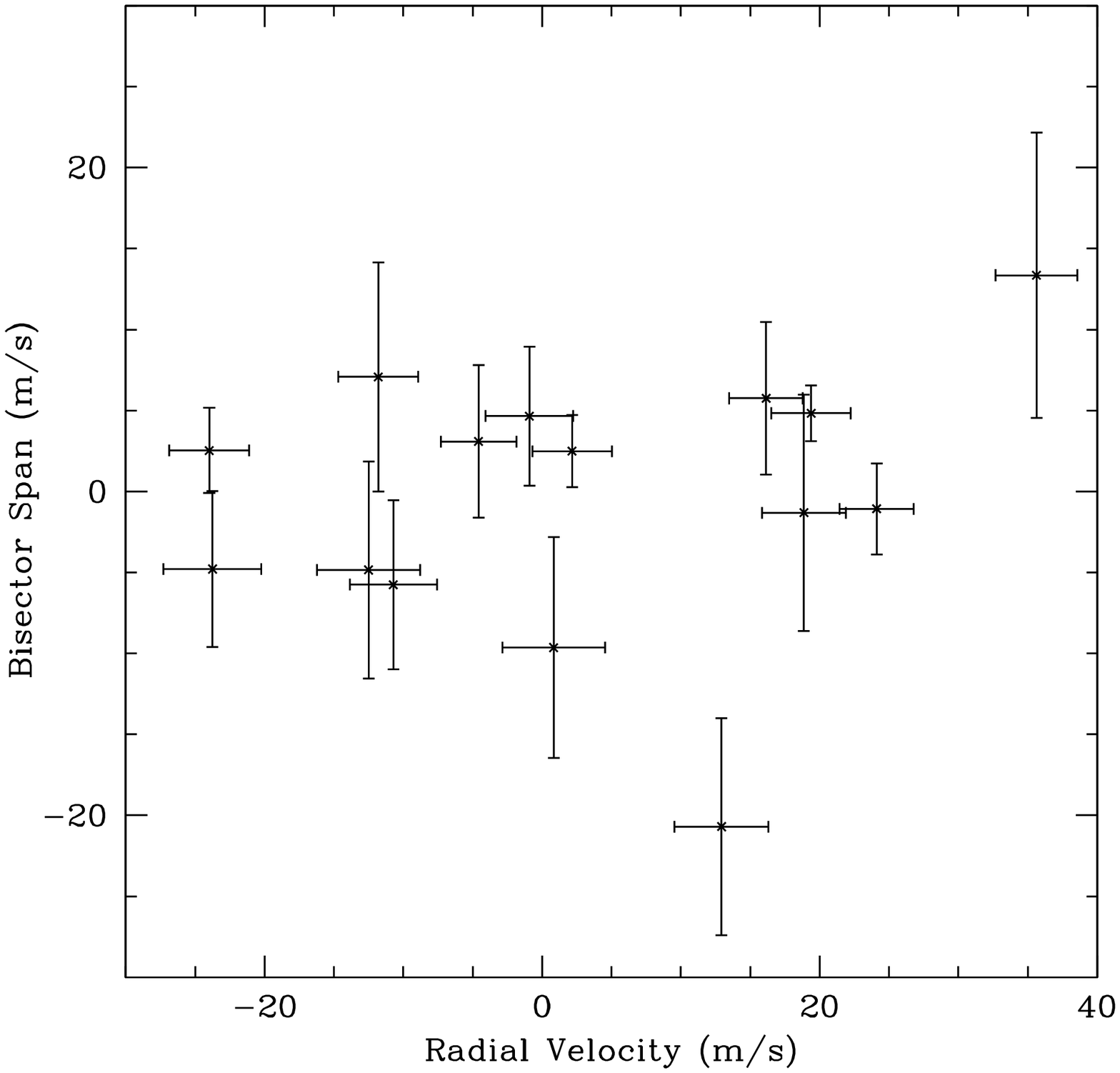}
\vspace{.1in}
\caption{Bisector spans for the APF RV spectra plotted against the RV values.  No correlation between the quantities is seen that would indicate that the RV signal is caused by line distortions from an unresolved eclipsing binary instead of true reflex motion of the star KELT-11.}
\label{fig:BIS} 
\end{figure}

We calculate bisector spans (BS) for each APF spectra following the prescription of \citet{Fulton:2015}. We cross-correlate each spectrum with a synthetic stellar template interpolated from the \citet{Coelho:2014} grid of stellar atmosphere models to match the adopted stellar parameters. We restrict the BS analysis to the 15 echelle orders spanning 4260 to 5000 \AA\ in order to avoid iodine contamination, telluric lines, and reduce the effect of telescope guiding errors on the instrumental PSF. The blue orders were used where the seeing is generally worse and the slit is more uniformly illuminated. This reduces false bisector variations caused by changes in the slit illumination. The BS is a measure of the line asymmetry calculated as the difference in the midpoint of line segments drawn between the 65th and 95th percentile levels of the CCF. The reported BSs and uncertainties are the mean and standard deviation on the mean over the 15 spectral orders analyzed.  The BS values are included in Table \ref{tab:RVs}, and are plotted against the RV values in Figure \ref{fig:BIS}.

\subsection{High-Contrast Imaging}
\label{sec:AO}
    
\begin{figure}[!ht]
\centering\epsfig{file=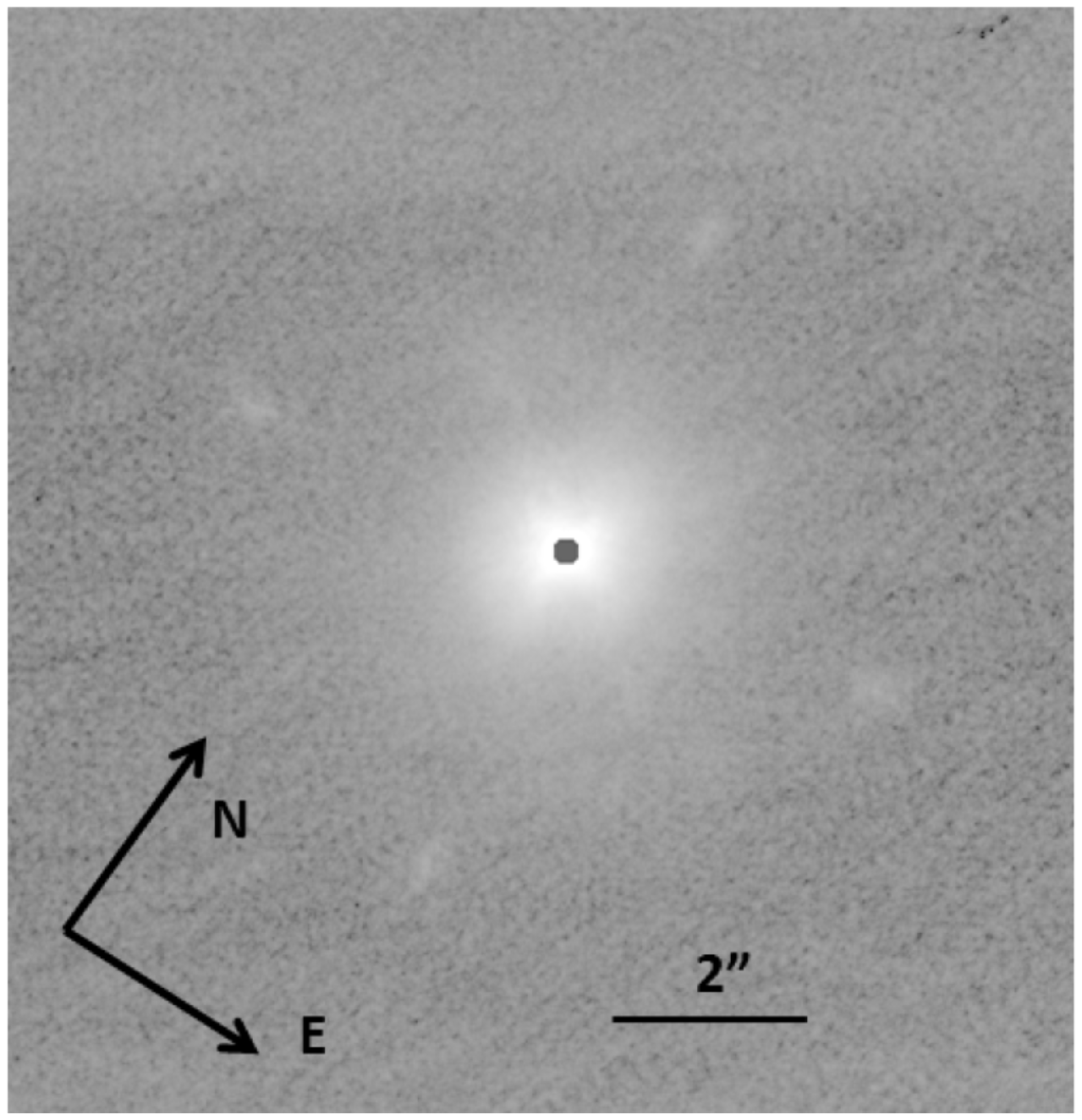,clip=,width=0.99\linewidth}
\caption{Palomar AO image of KELT-11. The four spots oriented cardinally with the target are a result of the alignment procedure and are not astrophysical.}
\label{fig:AO}
\end{figure}

\begin{figure}
  \centering \includegraphics[width=0.98\columnwidth]{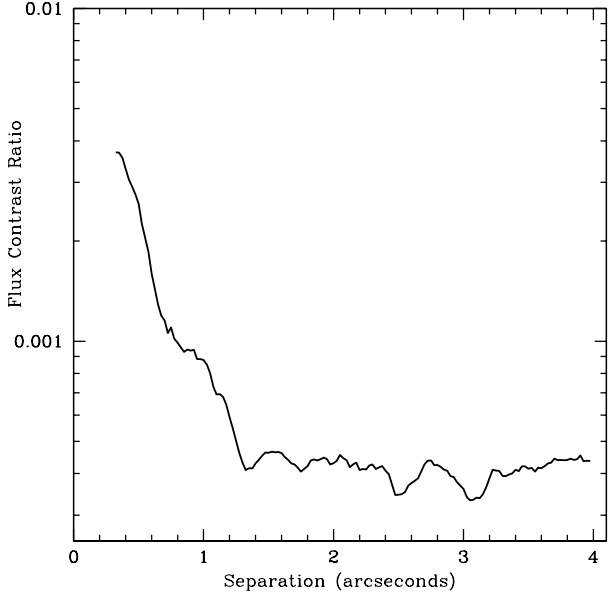}
  \vspace{.1in}
  \caption{Contrast curve based on Palomar AO imaging.  After removing systematics and the point-spread function of the target, no nearby stars are seen within an annulus of radii $\sim1.2\arcsec$ to $4\arcsec$ down to a flux ratio of $< 4 \times 10^{-4}$.}
  \label{fig:Contrast}
\end{figure}

We observed KELT-11 with Adaptive Optics (AO) imaging to identify any faint, nearby companions to the host star.  The observations were performed at Palomar Observatory's Hale telescope on 2015 February 7 using the Stellar Double Coronagraph \citep{Bottom:2015} operating with a ring-apodized vortex.  Observations were taken in the $K$-short band, obtaining 60 exposures at 29.7 sec each. Figure \ref{fig:AO} shows the image of the star, and Figure \ref{fig:Contrast} shows the associated contrast curve.  No astrophysical sources are seen near KELT-11.  At separations of 1.5 to 4 arcsec, which at the distance of KELT-11 corresponds to a projected separation of about 14 to 98 AU, we can exclude companions with flux ratios  of $4 \times 10^{-4}$ compared to KELT-11, which includes stars down to mid-M type dwarfs.

\begin{table*}
 \scriptsize
\centering
\setlength\tabcolsep{1.5pt}
\caption{Median values and 68\% confidence interval for the physical and orbital parameters of the KELT-11 system}
  \label{tab:KELT-11b_global_fit_properties}
  \begin{tabular}{lccccc}
  \hline
  \hline
   Parameter & Units & \textbf{Adopted Value} & Value & Value & Value \\
   & & \textbf{(YY circular; $e$=0 fixed)} & (YY eccentric) & (Torres circular; $e$=0 fixed) & (Torres eccentric)\\
 \hline
Stellar Parameters & & & & &\\
                               ~~~$M_{*}$\dotfill &Mass (\msun)\dotfill & $1.438_{-0.052}^{+0.061}$&$1.463_{-0.056}^{+0.064}$&$1.347_{-0.068}^{+0.072}$&$1.364_{-0.068}^{+0.070}$\\
                             ~~~$R_{*}$\dotfill &Radius (\rsun)\dotfill & $2.72_{-0.17}^{+0.21}$&$2.81_{-0.19}^{+0.21}$&$2.78_{-0.22}^{+0.24}$&$2.89\pm0.22$\\
                         ~~~$L_{*}$\dotfill &Luminosity (\lsun)\dotfill & $5.55_{-0.69}^{+0.85}$&$5.90_{-0.79}^{+0.86}$&$5.86_{-0.90}^{+1.1}$&$6.28_{-0.92}^{+1.0}$\\
                             ~~~$\rho_*$\dotfill &Density (cgs)\dotfill & $0.101\pm0.017$&$0.093_{-0.015}^{+0.018}$&$0.088_{-0.017}^{+0.022}$&$0.080_{-0.014}^{+0.018}$\\
                  ~~~\logg\ \dotfill &Surface gravity (cgs)\dotfill & $3.727_{-0.046}^{+0.040}$&$3.706\pm0.044$&$3.678_{-0.058}^{+0.060}$&$3.653_{-0.053}^{+0.056}$\\
                  ~~~$\teff$\dotfill &Effective temperature (K)\dotfill & $5370_{-50}^{+51}$&$5367\pm50$&$5386\pm49$&$5385_{-50}^{+49}$\\
                                 ~~~$\feh$\dotfill &Metallicity\dotfill & $0.180\pm0.075$&$0.173_{-0.077}^{+0.076}$&$0.194\pm0.079$&$0.191\pm0.079$\\
\hline
 Planet Parameters & & & & & \\
                                   ~~~$e$\dotfill &Eccentricity\dotfill & --- &$0.066_{-0.046}^{+0.059}$& --- &$0.087_{-0.058}^{+0.076}$\\
        ~~~$\omega_*$\dotfill &Argument of periastron (degrees)\dotfill & --- &$97_{-65}^{+82}$& --- &$86_{-46}^{+69}$\\
                                  ~~~$P$\dotfill &Period (days)\dotfill & $4.736529_{-0.000059}^{+0.000068}$&$4.736525_{-0.000057}^{+0.000067}$&$4.736528_{-0.000062}^{+0.000069}$&$4.736519_{-0.000058}^{+0.000063}$\\
                           ~~~$a$\dotfill &Semi-major axis (AU)\dotfill & $0.06229_{-0.00076}^{+0.00088}$&$0.06264_{-0.00081}^{+0.00089}$&$0.0609_{-0.0010}^{+0.0011}$&$0.0612\pm0.0010$\\
                                 ~~~$M_{P}$\dotfill &Mass (\mj)\dotfill & $0.195_{-0.018}^{+0.019}$&$0.199\pm0.021$&$0.187\pm0.018$&$0.190_{-0.020}^{+0.021}$\\
                               ~~~$R_{P}$\dotfill &Radius (\rj)\dotfill & $1.37_{-0.12}^{+0.15}$&$1.41_{-0.12}^{+0.14}$&$1.42_{-0.15}^{+0.17}$&$1.46_{-0.14}^{+0.15}$\\
                           ~~~$\rho_{P}$\dotfill &Density (cgs)\dotfill & $0.093_{-0.024}^{+0.028}$&$0.088_{-0.021}^{+0.027}$&$0.081_{-0.022}^{+0.030}$&$0.076_{-0.019}^{+0.025}$\\
                      ~~~$\log{g_{P}}$\dotfill &Surface gravity\dotfill & $2.407_{-0.086}^{+0.080}$&$2.392_{-0.083}^{+0.081}$&$2.360_{-0.096}^{+0.093}$&$2.347_{-0.087}^{+0.085}$\\
               ~~~$T_{eq}$\dotfill &Equilibrium temperature (K)\dotfill & $1712_{-46}^{+51}$&$1733_{-50}^{+49}$&$1755\pm64$&$1782\pm60$\\
                           ~~~$\Theta^a$\dotfill &Safronov number\dotfill & $0.0123_{-0.0015}^{+0.0016}$&$0.0120_{-0.0016}^{+0.0017}$&$0.0119_{-0.0016}^{+0.0017}$&$0.0117_{-0.0015}^{+0.0017}$\\
                   ~~~$\fave$\dotfill &Incident flux (\fluxcgs)\dotfill & $1.95_{-0.20}^{+0.24}$&$2.04_{-0.22}^{+0.23}$&$2.15_{-0.30}^{+0.33}$&$2.26_{-0.28}^{+0.30}$\\
 \hline
 RV Parameters & & & & & \\
       ~~~$T_C$\dotfill &Time of inferior conjunction (\bjdtdb)\dotfill & $2457061.9098_{-0.0024}^{+0.0026}$&$2457061.9098_{-0.0025}^{+0.0027}$&$2457061.9101_{-0.0025}^{+0.0027}$&$2457061.9101_{-0.0024}^{+0.0026}$\\
               ~~~$T_{P}$\dotfill &Time of periastron (\bjdtdb)\dotfill & --- &$2457061.99_{-0.74}^{+1.0}$& --- &$2457061.89_{-0.48}^{+0.84}$\\
                        ~~~$K$\dotfill &RV semi-amplitude (m/s)\dotfill & $18.5\pm1.7$&$18.7\pm1.9$&$18.5_{-1.6}^{+1.7}$&$18.8_{-1.9}^{+2.0}$\\
                    ~~~$M_P\sin{i}$\dotfill &Minimum mass (\mj)\dotfill & $0.195\pm0.018$&$0.198\pm0.021$&$0.186_{-0.017}^{+0.018}$&$0.190_{-0.020}^{+0.021}$\\
                           ~~~$M_{P}/M_{*}$\dotfill &Mass ratio\dotfill & $0.000129\pm0.000012$&$0.000130\pm0.000013$&$0.000133\pm0.000012$&$0.000133_{-0.000013}^{+0.000014}$\\
                       ~~~$u$\dotfill &RM linear limb darkening\dotfill & $0.7061_{-0.0073}^{+0.0071}$&$0.7058\pm0.0072$&$0.7046_{-0.0073}^{+0.0072}$&$0.7044_{-0.0074}^{+0.0072}$\\
                                 ~~~$\gamma_{APF}$\dotfill &m/s\dotfill & $1.9\pm2.4$&$1.8_{-2.7}^{+2.6}$&$1.9\pm2.3$&$1.8_{-2.6}^{+2.7}$\\
                                ~~~$\gamma_{KECK}$\dotfill &m/s\dotfill & $-1.8\pm2.4$&$-1.6\pm2.7$&$-1.9\pm2.4$&$-1.7\pm2.7$\\
                  ~~~$\dot{\gamma}$\dotfill &RV slope (m/s/day)\dotfill & $-0.0060\pm0.0015$&$-0.0059\pm0.0017$&$-0.0060\pm0.0015$&$-0.0060\pm0.0017$\\
                                         ~~~$\ecosw$\dotfill & \dotfill & --- &$-0.004_{-0.051}^{+0.049}$& --- &$0.000_{-0.050}^{+0.064}$\\
                                         ~~~$\esinw$\dotfill & \dotfill & --- &$0.031_{-0.042}^{+0.068}$& --- &$0.058_{-0.060}^{+0.084}$\\
 \hline
 \hline
 \end{tabular}
\begin{flushleft}
    \hspace{0.5in}
  \footnotesize $^a$ \citet{Hansen:2007} \\
  \end{flushleft}
\end{table*}

\begin{table*}
\scriptsize
 \centering
\setlength\tabcolsep{1.5pt}
\caption{Median values and 68\% confidence interval for the physical and orbital parameters of the KELT-11 system (continued)}
  \label{tab:KELT-11b_global_fit_properties_p2}
  \begin{tabular}{lccccc}
  \hline
  \hline
   Parameter & Units & \textbf{Adopted Value} & Value & Value & Value \\
   & & \textbf{(YY circular; $e$=0 fixed)} & (YY eccentric) & (Torres circular; $e$=0 fixed) & (Torres eccentric)\\
 \hline
 \hline
 Primary Transit & & & & & \\
~~~$R_{P}/R_{*}$\dotfill &Radius of the planet in stellar radii\dotfill & $0.0519\pm0.0026$&$0.0516\pm0.0026$&$0.0526\pm0.0027$&$0.0520\pm0.0026$\\
           ~~~$a/R_*$\dotfill &Semi-major axis in stellar radii\dotfill & $4.93_{-0.29}^{+0.26}$&$4.80_{-0.27}^{+0.28}$&$4.71_{-0.32}^{+0.36}$&$4.56_{-0.29}^{+0.32}$\\
                          ~~~$i$\dotfill &Inclination (degrees)\dotfill & $85.8_{-1.8}^{+2.4}$&$86.4_{-2.2}^{+2.3}$&$84.4_{-1.8}^{+2.6}$&$85.4_{-2.3}^{+2.9}$\\
                               ~~~$b$\dotfill &Impact parameter\dotfill & $0.36_{-0.20}^{+0.13}$&$0.29_{-0.19}^{+0.17}$&$0.46_{-0.19}^{+0.11}$&$0.34_{-0.22}^{+0.17}$\\
                             ~~~$\delta$\dotfill &Transit depth\dotfill & $0.00269_{-0.00026}^{+0.00028}$&$0.00267_{-0.00026}^{+0.00027}$&$0.00276_{-0.00028}^{+0.00029}$&$0.00271_{-0.00027}^{+0.00028}$\\
                    ~~~$T_{FWHM}$\dotfill &FWHM duration (days)\dotfill & $0.2874\pm0.0048$&$0.2873\pm0.0047$&$0.2876_{-0.0048}^{+0.0049}$&$0.2877_{-0.0045}^{+0.0048}$\\\
              ~~~$\tau$\dotfill &Ingress/egress duration (days)\dotfill & $0.0173_{-0.0021}^{+0.0029}$&$0.0164_{-0.0016}^{+0.0029}$&$0.0193_{-0.0032}^{+0.0038}$&$0.0172_{-0.0021}^{+0.0039}$\\
                     ~~~$T_{14}$\dotfill &Total duration (days)\dotfill & $0.3051_{-0.0051}^{+0.0053}$&$0.3043_{-0.0052}^{+0.0053}$&$0.3072_{-0.0053}^{+0.0059}$&$0.3056_{-0.0052}^{+0.0058}$\\
   ~~~$P_{T}$\dotfill &A priori non-grazing transit probability\dotfill & $0.1925_{-0.0096}^{+0.012}$&$0.205_{-0.018}^{+0.024}$&$0.201\pm0.014$&$0.221_{-0.024}^{+0.033}$\\
             ~~~$P_{T,G}$\dotfill &A priori transit probability\dotfill & $0.214_{-0.011}^{+0.013}$&$0.228_{-0.020}^{+0.027}$&$0.224_{-0.016}^{+0.017}$&$0.246_{-0.027}^{+0.037}$\\
               ~~~$T_{C,0}$\dotfill &Mid-transit time (\bjdtdb)\dotfill & $2457024.0176_{-0.0028}^{+0.0030}$&$2457024.0175_{-0.0029}^{+0.0030}$&$2457024.0179_{-0.0029}^{+0.0031}$&$2457024.0179_{-0.0027}^{+0.0030}$\\
               ~~~$T_{C,1}$\dotfill &Mid-transit time (\bjdtdb)\dotfill & $2457061.9098_{-0.0024}^{+0.0026}$&$2457061.9098_{-0.0025}^{+0.0027}$&$2457061.9101_{-0.0025}^{+0.0027}$&$2457061.9101_{-0.0024}^{+0.0026}$\\
               ~~~$T_{C,2}$\dotfill &Mid-transit time (\bjdtdb)\dotfill & $2457061.9098_{-0.0024}^{+0.0026}$&$2457061.9098_{-0.0025}^{+0.0027}$&$2457061.9101_{-0.0025}^{+0.0027}$&$2457061.9101_{-0.0024}^{+0.0026}$\\
               ~~~$T_{C,3}$\dotfill &Mid-transit time (\bjdtdb)\dotfill & $2457061.9098_{-0.0024}^{+0.0026}$&$2457061.9098_{-0.0025}^{+0.0027}$&$2457061.9101_{-0.0025}^{+0.0027}$&$2457061.9101_{-0.0024}^{+0.0026}$\\
               ~~~$T_{C,4}$\dotfill &Mid-transit time (\bjdtdb)\dotfill & $2457090.3290_{-0.0022}^{+0.0024}$&$2457090.3289_{-0.0022}^{+0.0024}$&$2457090.3293_{-0.0023}^{+0.0025}$&$2457090.3292_{-0.0022}^{+0.0024}$\\
               ~~~$T_{C,5}$\dotfill &Mid-transit time (\bjdtdb)\dotfill & $2457095.0655_{-0.0022}^{+0.0024}$&$2457095.0655_{-0.0022}^{+0.0024}$&$2457095.0658_{-0.0022}^{+0.0024}$&$2457095.0657_{-0.0022}^{+0.0024}$\\
               ~~~$T_{C,6}$\dotfill &Mid-transit time (\bjdtdb)\dotfill & $2457099.8021_{-0.0021}^{+0.0024}$&$2457099.8020_{-0.0022}^{+0.0024}$&$2457099.8024_{-0.0022}^{+0.0024}$&$2457099.8023_{-0.0021}^{+0.0024}$\\
               ~~~$T_{C,7}$\dotfill &Mid-transit time (\bjdtdb)\dotfill & $2457147.1674_{-0.0019}^{+0.0021}$&$2457147.1673_{-0.0020}^{+0.0021}$&$2457147.1677_{-0.0019}^{+0.0021}$&$2457147.1675_{-0.0019}^{+0.0021}$\\
               ~~~$T_{C,8}$\dotfill &Mid-transit time (\bjdtdb)\dotfill & $2457440.8323_{-0.0036}^{+0.0042}$&$2457440.8319_{-0.0036}^{+0.0041}$&$2457440.8325_{-0.0038}^{+0.0043}$&$2457440.8318_{-0.0037}^{+0.0040}$\\
                     ~~~$u_{1I}$\dotfill &Linear Limb-darkening\dotfill & $0.3466_{-0.0093}^{+0.0094}$&$0.3462_{-0.0092}^{+0.0094}$&$0.3437\pm0.0094$&$0.3434\pm0.0094$\\
                  ~~~$u_{2I}$\dotfill &Quadratic Limb-darkening\dotfill & $0.2510\pm0.0050$&$0.2511\pm0.0049$&$0.2533\pm0.0050$&$0.2536\pm0.0050$\\
                     ~~~$u_{1R}$\dotfill &Linear Limb-darkening\dotfill & $0.446\pm0.012$&$0.446\pm0.012$&$0.443\pm0.012$&$0.443\pm0.012$\\
                  ~~~$u_{2R}$\dotfill &Quadratic Limb-darkening\dotfill & $0.2374_{-0.0071}^{+0.0070}$&$0.2376\pm0.0070$&$0.2398_{-0.0071}^{+0.0070}$&$0.2401_{-0.0070}^{+0.0071}$\\
                ~~~$u_{1Sloang}$\dotfill &Linear Limb-darkening\dotfill & $0.695\pm0.017$&$0.695\pm0.017$&$0.692\pm0.017$&$0.692\pm0.017$\\
             ~~~$u_{2Sloang}$\dotfill &Quadratic Limb-darkening\dotfill & $0.117\pm0.013$&$0.117\pm0.013$&$0.120\pm0.013$&$0.120\pm0.013$\\
                ~~~$u_{1Sloani}$\dotfill &Linear Limb-darkening\dotfill & $0.371\pm0.010$&$0.3702_{-0.0100}^{+0.010}$&$0.368\pm0.010$&$0.367\pm0.010$\\
             ~~~$u_{2Sloani}$\dotfill &Quadratic Limb-darkening\dotfill & $0.2485_{-0.0055}^{+0.0054}$&$0.2486\pm0.0054$&$0.2508\pm0.0055$&$0.2511_{-0.0055}^{+0.0054}$\\
                ~~~$u_{1Sloanr}$\dotfill &Linear Limb-darkening\dotfill & $0.476\pm0.013$&$0.475\pm0.013$&$0.472\pm0.013$&$0.472\pm0.013$\\
             ~~~$u_{2Sloanr}$\dotfill &Quadratic Limb-darkening\dotfill & $0.2316_{-0.0078}^{+0.0077}$&$0.2318_{-0.0077}^{+0.0076}$&$0.2340\pm0.0077$&$0.2343\pm0.0077$\\
                ~~~$u_{1Sloanz}$\dotfill &Linear Limb-darkening\dotfill & $0.2976\pm0.0075$&$0.2974\pm0.0074$&$0.2952_{-0.0075}^{+0.0074}$&$0.2949\pm0.0075$\\
             ~~~$u_{2Sloanz}$\dotfill &Quadratic Limb-darkening\dotfill & $0.2549_{-0.0036}^{+0.0037}$& $0.2549_{-0.0036}^{+0.0035}$&$0.2568\pm0.0036$&$0.2570\pm0.0036$\\
\hline
Secondary Eclipse & & & & & \\
                  ~~~$T_{S}$\dotfill &Time of eclipse (\bjdtdb)\dotfill & $2457059.5415_{-0.0025}^{+0.0027}$&$2457064.27_{-0.16}^{+0.15}$&$2457059.5418_{-0.0025}^{+0.0027}$&$2457064.28_{-0.15}^{+0.19}$\\
                           ~~~$b_{S}$\dotfill &Impact parameter\dotfill & --- &$0.31_{-0.20}^{+0.16}$& --- &$0.39_{-0.24}^{+0.15}$\\
                  ~~~$T_{S,FWHM}$\dotfill &FWHM duration (days)\dotfill & --- &$0.303_{-0.022}^{+0.042}$& --- &$0.315_{-0.029}^{+0.057}$\\
            ~~~$\tau_S$\dotfill &Ingress/egress duration (days)\dotfill & --- &$0.0183_{-0.0024}^{+0.0029}$& --- &$0.0205_{-0.0031}^{+0.0039}$\\
                   ~~~$T_{S,14}$\dotfill &Total duration (days)\dotfill & --- &$0.322_{-0.023}^{+0.043}$& --- &$0.336_{-0.030}^{+0.058}$\\
   ~~~$P_{S}$\dotfill &A priori non-grazing eclipse probability\dotfill & --- &$0.1890_{-0.0069}^{+0.013}$& --- &$0.1934_{-0.0093}^{+0.016}$\\
             ~~~$P_{S,G}$\dotfill &A priori eclipse probability\dotfill & --- &$0.2095_{-0.0079}^{+0.014}$& --- &$0.215_{-0.011}^{+0.018}$\\
     \hline
 \hline
\end{tabular}
\end{table*}
    
\section{Analysis and Results}

\subsection{Stellar Parameters from Spectra}
\label{sec:spec_pars}

To obtain a thorough picture of the stellar parameters, we used multiple parameter extraction techniques with different spectra to obtain consistent and robust results.

We applied the Spectral Parameter Classification (SPC) \citep{Buchhave:2012} technique to the spectrum from TRES, with $\teff$, \logg, [m/H], and $v \sin I_*$ as free parameters. SPC cross-correlates an observed spectrum against a grid of synthetic spectra based on Kurucz atmospheric models \citep{Kurucz:1992}. The resulting parameters are: $\teff = 5390 \pm 50$ K, \logg $= 3.789 \pm 0.100$, [m/H] = $0.189 \pm 0.080$ and $v \sin I_* = 2.66 \pm 0.50$ km/s.  These parameters match the identification of KELT-11 from \citet{Houk:1999} as a G8/K0 IV spectral type.

We then independently analyzed the Keck and APF template spectra using a modified version of the \texttt{SpecMatch} pipeline \citep{Petigura:2015} as described in \citet{Fulton:2015b}. \texttt{SpecMatch} fits a grid of model stellar atmospheres \citep{Coelho:2014} to large regions of the observed spectrum to measure $\teff$, \logg, and [Fe/H].  Our only modification to the code is to employ the ExoPy differential-evolution Markov Chain Monte Carlo fitting engine \citep{Fulton:2013} in place of $\chi^2$ minimization to find the optimal solution. We calibrate the $\teff$ and \logg\ values produced by \texttt{SpecMatch} to show good agreement with the values of \citet{Valenti:2005} for the 353 stars in their sample that were observed on Keck. Values of \logg\ are calibrated against the stars from the \citet{Huber:2012} sample with asteroseismically measured \logg. This calibration is done separately for both APF and Keck spectra using our modified version of the pipeline. However, the faint \emph{Kepler} stars from the \citep{Huber:2012} sample have not been observed on APF so the \logg\ calibration is bootstrapped from stars that have been observed at both Keck and APF. Parameter uncertainties are determined by measuring the scatter of our calibrated parameters against the values form \citet{Valenti:2005} and \citet{Huber:2012}. 

The best fit spectroscopic parameters measured from all three reductions are listed in Table \ref{tab:stellar_params}.
 
\subsection{SED Analysis}
\label{sec:sed}
We estimate the distance and reddening to KELT-11
by fitting Kurucz (1979) stellar atmosphere models to the spectral
energy distribution (SED) from catalog broad-band photometry. 
The available catalog photometry spans the range from the GALEX FUV band at 0.16~$\mu$m to the WISE 22.8~$\mu$m band (see Table~1). 

We fix $T_{\rm eff}$, \logg\, and [Fe/H] to the TRES-derived values from \S \ref{sec:spec_pars}, and leave the distance ($d$) and reddening ($A_V$) as free parameters. We restricted $A_V$ to be at most 0.11 mag, as determined from the maximum line-of-sight extinction from the dust maps of \citet{Schlegel:1998}.  We find best-fit parameters of $A_V = 0.09^{+0.02}_{-0.05}$ mag and $d= 98\pm5$ pc. This agrees quite well with, and is consistent with, the distance of 102$\pm$9 pc from the Hipparcos parallax. The reduced $\chi^2$ of the fit is 7.0 when including the GALEX FUV and NUV photometry, but 1.9 when excluding the GALEX photometry. This reflects the apparent excess in the GALEX bands, suggestive of chromospheric activity.  While it may be surprising to see such activity, similarly chromospherically active subgiants are rare but are seen in previous surveys of activity among subgiants \citep{Jenkins:2011}.  We looked for evidence of chromospheric activity in the Ca H and K lines of the APF spectra, and do not find any.  Since the UV excess apparent in the SED is not corroborated by the Ca II H and K spectroscopy, we do not use the UV flux excess for any other aspects of our analysis or interpretations. 

One possible cause of excess chromospheric activity can be spin-up of the host star from a dynamically interacting planet.  We can explore that possibility in this case since we have a spectroscopic $v \sin I_*$ measurement.  Assuming the star and planet are not highly misaligned, and therefore that $\sin I_* \sim 1$, we have a rotational period for the star of roughly 52 days.  That is much longer than the orbital period of the planet, and so we do not find evidence for an unusually high rotational velocity due to planet spin-up of the star.

Using the Hipparcos distance and the spectroscopic $T_{\rm eff}$ we can measure the stellar radius independently by summing the SED flux and solving the Stefan-Boltzmann relation. The summed bolometric flux received at Earth is $1.85\pm0.06\times10^{-8}$ erg s$^{-1}$ cm$^{-2}$, and the resulting stellar radius is 2.84$\pm$0.28 R$_\odot$, which we use as a prior on the global analysis of the transit (see \S \ref{sec:exofast}).

\begin{figure}
  \centering \includegraphics[width=0.75\columnwidth, angle=90]{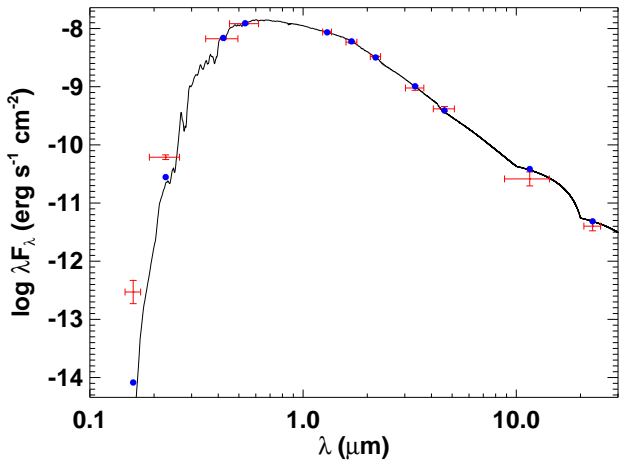}
  \caption{Measured and best fitting spectral energy distribution (SED) for KELT-11 from ultraviolet through mid-infrared. The red error bars indicate measurements of the flux of KELT-11 in ultraviolet, optical, near-infrared  and mid-infrared passbands as listed in Table \ref{tab:Host_Lit_Props}. The vertical bars are the 1$\sigma$ photometric uncertainties, whereas the horizontal error bars are the effective widths of the passbands. The solid curve is the best-fitting theoretical SED from the models of \citet{Kurucz:1992}, assuming stellar parameters $\teff$ , $\log{g_*}$ and $\feh$ fixed at the adopted values in Table \ref{tab:KELT-11b_global_fit_properties} and \ref{tab:KELT-11b_global_fit_properties_p2}, with the extinction and distance (A$_v$ and $d$) allowed to vary. The blue dots are the predicted passband-integrated fluxes of the best-fitting theoretical SED corresponding to our observed photometric bands. (A colour version of this figure is available in the online journal.)}
  \label{fig:SED_figure}
\end{figure}

\subsection{Evolutionary Analysis}
\label{sec:evol}
\begin{figure}
\centering
\includegraphics[width=0.75\columnwidth,angle=90]{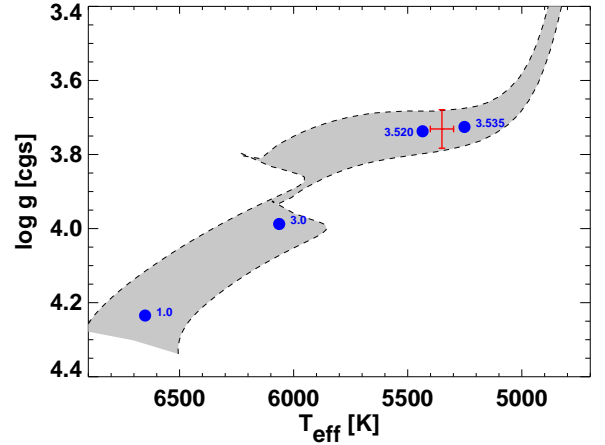}
\caption{A theoretical H-R diagram for KELT-11b using the Yonsei-Yale stellar evolution models \citep{Demarque:2004}. The final global fit values are shown by the red cross. The blue points mark various ages along the tracks. The gray shaded region is a 1$\sigma$ uncertainty on [Fe/H] and M$_{\star}$ from the global modeling.}
\label{fig:hrd}
\end{figure}

We estimate the age of KELT-11 by fitting Yonsei-Yale isochrones to the values of $T_{\rm eff}$, \logg, and [Fe/H] given in Table \ref{tab:KELT-11b_global_fit_properties}. We fix the stellar mass to the value of 1.438 M$_\odot$ as listed in Table \ref{tab:KELT-11b_global_fit_properties}. Our best fit stellar parameters indicate that the star happens to fall on a rapid part of the evolutionary track, namely the so-called ``Hertzsprung gap" prior to the star's ascent up the red giant branch (see Figure \ref{fig:hrd}).  This permits a relatively precise, although model-dependent, age estimate, and we quote an age range of 3.52--3.54 Gyr that spans the uncertainty in $T_{\rm eff}$ and \logg.

\subsection{UVW Space Motion}

We examined the three-dimensional space motion of KELT-11 to determine whether its kinematics match that of one of the main stellar populations of the Galaxy.  The proper motion is $-78.30 \pm 0.95$ mas/year in right ascension, and $-77.61 \pm 0.68$ mas/year in declination \citep{Leeuwen:2007}.  The values for the absolute heliocentric RV as measured by the spectroscopic observations from Keck and APF (see \S \ref{sec:keck} and \S \ref{sec:apf}) are mutually consistent, and we adopt the value from the Keck results, which is $35.0 \pm 0.1$ \kms.  The distance to the system as derived in \S \ref{sec:sed} is $98 \pm 5$ pc.  These values transform to U, V, W space motions with respect to the local standard of rest (LSR) of -9.6 $\pm$ 1.2, -46.2 $\pm$ 3.0, and -8.4 $\pm$ 3.4 \kms, respectively.  When calculating the UVW motion, we have accounted for the peculiar velocity of the Sun with respect to the LSR, using U = 8.5 \kms, V = 13.38 \kms, and W = 6.49 \kms from \citet{LSR:2011}.  These values are consistent with that of a fast-moving thin disk star \citep{Bensby:2003}.

\subsection{EXOFAST Global Fit}
\label{sec:exofast}
Using a modified version of the IDL exoplanet fitting package, EXOFAST \citep{Eastman:2013}, we perform a simulataneous Markov Chain Monte Carlo (MCMC) analysis on the KECK HIRES and APF RV measurements, and the follow-up photometric observations by the KELT Follow-up Network.  See \citet{Siverd:2012} for a more complete description of the global modeling.  For KELT-11b, we use either the Yonsei-Yale stellar evolution models \citep{Demarque:2004} or the Torres relations \citep{Torres:2010} to constrain M$_{\star}$ and R$_{\star}$. Our SED analysis (\S \ref{sec:sed}) of the catalog photometry of KELT-11 yielded an observed bolometric flux from the star of $1.82\times10^{-8}$ erg s$^{-1}$ cm$^{-2}$. 
We use the stellar radius of $\mathrm{R}_*=2.8\pm0.3$\rsun\, as a prior in the global fit. From the KELT discovery data and follow-up photometric and spectroscopic observations, we set a prior on the host star's effective temperature (\teff = 5391 $\pm$ 50 K) and metallicity (\feh = 0.189 $\pm$ 0.08). The raw light curve and detrending parameters (see Table \ref{tab:followup_lcs} for detrending parameters of each dataset) are inputs into the final global fit using EXOFAST. For KELT-11b, we ran four separate EXOFAST fits using either the Torres relations or Yonsei-Yale stellar evolution models. For each stellar model, we ran one fit where the eccentricity of the planet's orbit was a free parameter and another where it was fixed at an eccentricity of zero. The results of these fits are shown in Tables \ref{tab:KELT-11b_global_fit_properties} and \ref{tab:KELT-11b_global_fit_properties_p2}.  In Table \ref{tab:KELT-11b_global_fit_properties_p2}, the quantities $T_{C,1}$, $T_{C,2}$, etc. refer to the determination of the transit center time for each of the follow-up light curves used in the model, shown in Figure \ref{fig:All_Lightcurve}.  Because all but one of our follow-up photometry light curves are partial transits, we do not fit for transit timing variations as part of the global fit. The secondary eclipse parameters in all cases are are not directly measured, but are rather based on the global fit.  All four global fits are consistent with each other to within 2$\sigma$. For our overall analysis and discussion, we adopt the Yonsei-Yale circular fit parameters. 

Our stellar parameter priors strongly favor young, pre-main sequence stars, and also implies a mass prior strongly peaked at 1 M$_{\odot}$ with a long tail to higher masses. However, our global fit, due to the stellar density constraint from the transit, excludes the peak of this prior with high confidence, enhancing the high mass tail. The prior at the location of the ultimately inferred mass is biased slightly toward lower mass stars relative to a flat, ``uninformative" prior. We believe this is justified because generally lower-mass stars with temperatures and surface gravities consistent with those measured for KELT-11 are favored because both higher mass stars are rarer and spend less time in this region of the HR diagram.

We find a negative slope in the RV observations of $-0.006\pm0.0015$ m s$^{-1}$ day$^{-1}$.  That is most likely a result of an outer massive companion in orbit around KELT-11.  Since there is no evidence of significant curvature in the slope, the companion must have an orbital period longer than about 20 years.  The companion could be a large planet or a brown dwarf, but it likely cannot be stellar mass, since there are no signs of an additional object in the spectral analysis.  As noted in \S \ref{sec:AO}, there are no indications of a companion in the AO observations, although those are most sensitive at wide separations from the star.

\subsection{Irradiation History}

We find that KELT-11b is a highly inflated planet, 
joining the ranks of other gas giant planets that
manifest radii much larger than predicted by standard models of irradiated objects with Jovian masses, which do not invoke additional sources of energy deposition in the interior of the planet.
Several authors \citep[e.g.,][]{Demory:2011} have suggested
an empirical insolation threshold ($\approx 2 \times 10^8$
erg s$^{-1}$ cm$^{-2}$) above which gas giants exhibit increasing
amounts of radius inflation. KELT-11b clearly lies above
this threshold, with a current estimated insolation of
$1.95^{+0.24}_{-0.20} \times 10^9$ erg s$^{-1}$ cm$^{-2}$
and therefore its currently
large inflated radius is not surprising. At the same time,
the KELT-11 host star is found to currently be in a very
rapid state of evolution, such that its radius is rapidly expanding
as the star crosses the Hertzsprung gap toward
the red giant branch. This means that the star's surface
is rapidly encroaching on the planet, which presumably
is rapidly driving up the planet's insolation and also the
rate of any tidal interactions between the planet and the star.

Therefore it is interesting to consider whether KELT-11b's incident 
radiation from its host 
star has been below the empirical radius inflation threshold
in the past. If KELT-11b's insolation only recently
exceeded the inflation threshold, the system could then
serve as an empirical testbed for the different timescales
predicted by different inflation mechanisms \citep[see, e.g.,][]{Assef:2009,Spiegel:2012}. 

\begin{figure}
\centering
\includegraphics[width=0.75\columnwidth,angle=90]{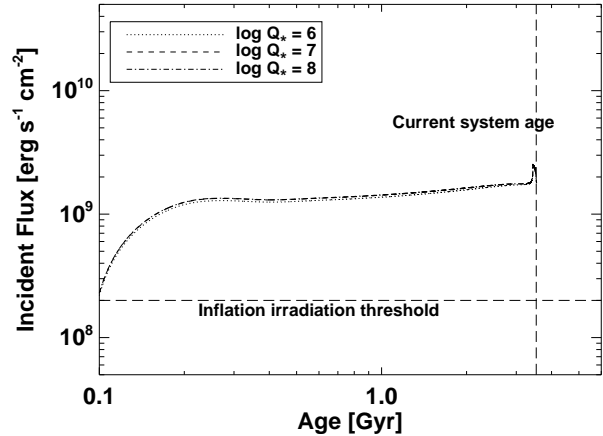}
\caption{A history of the inflation irradiation for KELT-11 for test values of $\log Q'_\star$ from 6 to 8. The horizontal dashed line indicates the insolation threshold determined by \citet{Demory:2011}}
\label{fig:evol}
\end{figure}

To investigate this, we follow \citet{Penev:2014} to simulate the past and future evolution of
the star-planet system, using the measured parameters
listed in Tables \ref{tab:KELT-11b_global_fit_properties} and \ref{tab:KELT-11b_global_fit_properties_p2} as the present-day boundary
conditions. This analysis is not intended to examine
any type of planet-planet or planet-disk migration effects.
Rather, it is a way to investigate the change in
insolation of the planet over time due to the changing luminosity
of the star and changing star-planet separation. 
We include the evolution of the star, assumed to follow the
Yonsei-Yale stellar model with mass and metallicity as
in Table \ref{tab:KELT-11b_global_fit_properties}. For simplicity we assume that the stellar
rotation is negligible and treat the star as a solid body.
We also assume a circular orbit aligned with the stellar
equator throughout the full analysis. The results of our
simulations are shown in Figure \ref{fig:evol}. We tested a range
of values for the tidal quality factor of the star $Q'_\star$, from
$\log Q'_\star = 6$ to $\log Q'_\star = 8$ (assuming a constant phase lag
between the tidal bulge and the star-planet direction).
$Q'_\star$ is defined as the tidal quality factor divided by the
Love number (i.e., $Q'_\star = Q_\star /k_2$). 

We find that in all cases the
planet has always received more than enough flux from
its host to keep the planet irradiated beyond the insolation
threshold identified by \citet{Demory:2011}.

\subsection{False Positive Analysis}

We examine several lines of evidence to explore whether the combination of observations of this system are caused by a standard false positive, such as a blended eclipsing binary.  First, we verify that all photometric observations of the transit are consistent with achromatic transit depths (see \S \ref{sec:Follow-up_Photometry}), since wavelength-dependent transit depths at the level detectable by our photometry would indicate that this is a blended eclipsing binary.  We then check that in the KELT-11 spectra there are no signs of a second stellar spectrum blended with the spectrum of the host star (see \S \ref{sec:rvs}), and we also see no faint companions blended with KELT-11b in the AO observations (\S \ref{sec:AO}).  We also find no correlation between the bisector spans and the measured RVs (see Figure \ref{fig:BIS}).  The stellar surface gravity ($\log (g) = 3.727^{+0.040}_{-0.046}$) derived from the transits in the global fit, with no spectroscopic \logg prior imposed, is consistent with the three spectroscopically derived \logg\ values (see Table \ref{tab:stellar_params}) within $\approx1\sigma$.  We therefore conclude that KELT-11b is a bona fide planet.

\subsection{The Mass of KELT-11}

We find that the KELT-11 host star is located in a relatively sparsely populated part of the Hertzsprung-Russell (H-R) diagram, namely the "Hertzsprung gap".  This is the part of H-R diagram where relatively massive stars ($M_* \ga 1.5 M_\odot$) have finished core hydrogen fusion, but have yet to initiate hydrogen shell fusion, i.e., have yet to reach the giant branch.  In other words, this region is typically populated by relatively massive subgiants.  This part of the diagram is sparsely populated because massive stars spend a very small amount of time in this phase of their evolution relative to the duration of their hydrogen main sequence or giant branch evolutionary states, and because massive stars are intrinsically more rare than less massive stars.  It is precisely the short duration of this massive star's subgiant phase that allows us to put such a precise (albeit model-dependent) constraint on the age of the star.

However, the rapid evolution through this part of the H-R diagram and the relative paucity of massive stars also implies that it is {\it a priori} unlikely that we would have found a transiting planet orbiting such a star.  We therefore must be more diligent than usual in justifying our claims of the estimated parameters of the host star. The parameter that most strongly controls the duration of the star's time during the subgiant phase is the stellar mass.  Therefore, we concentrate our subsequent discussion on our confidence in the accuracy of our estimate of the stellar mass, and in particular our inference that the star is relatively massive ($M_* \ga 1.3~M_\odot$).

The fiducal parameters (including mass) we adopt for KELT-11 are obtained using a global fit as described in \S \ref{sec:exofast}. Along with the photometric and RV data, this fit used priors on the stellar effective temperature and metallicity from high-resolution spectroscopy, and stellar radius as determined from the SED and Hipparcos parallax (see \S \ref{sec:sed}).  For our fiducial adopted values, we also used constraints from the YY evolutianary models and assumed zero eccentricity.  We also explored the parameters inferred assuming eccentric orbits, as well as using the \citet{Torres:2010} empirical relations for the mass and radius of the host star as a function of the stellar effective temperature, surface gravity, and metallicity.

 Our procedure, or similar procedures (see \citealt{Holman:2007,Sozzetti:2007}), are commonly used to estimate the masses and radii of transiting planet hosts and their planets. However, it is worth recalling that a single-lined spectroscopic binary with only a primary eclipse does not yield a complete solution to the system.  Indeed, as pointed out by \citet{Seager:2003}, the only parameter that is directly measured from the photometry and RVs of such a system is the parameter $a/R_*$, which can be related to the stellar density with the (typically reasonable) assumption that $R_p \ll R_*$ and $M_p \ll M_*$ (and a measurement or constraint on the eccentricity).  Therefore, there is a one-parameter degeneracy between $M_*$ and $R_*$ for such systems, unless one includes additional direct observables of the system, or invokes model isochrones or empirical relations based on other stars. Using the latter may have some pitfalls, as described below.

For stars with properties that are not too different from the sun, we can expect that both the YY isochrones and \citet{Torres:2010} relations should be fairly reliable.  However, it is worth pointing out that it is known that sometimes these do not agree.  In particular, \citet{Collins:2014} demonstrated that the parameters inferred for the host star of KELT-6b, which happens to be relatively metal poor ([Fe/H]$\sim -0.3$), are significantly different when obtained using the YY constraints relative to those obtained using the \citet{Torres:2010} relations.  This is likely because the YY isochrones and/or the \citet{Torres:2010} relations are not very accurate at low metallicity.  It is also worth noting that only a small subset of the stars used to calibrate the \citet{Torres:2010} relations have metallicity measurements.

In the case of the KELT-11, we are fortunate to have another constraint on the property of the host star, specifically the host star radius.  However, there are also some unusual difficulties with this system. Because the parallax measurement is relatively imprecise ($\sim$ 10\%), this radius constraint is relatively poor.  Furthermore, the constraint on the stellar density measured directly from the light curve and the radial velocities is also relatively poor, due to the shallow depth and long duration, and the relatively weak constraint on the eccentricity.  Therefore, our final estimate of the mass and radius of KELT-11 is strongly influenced by our prior from the YY models or the \citet{Torres:2010} relations.  This is problematic because the YY models have not been well-calibrated for massive subdwarfs (because there are so few of them with direct mass measurements), and because there are no stars in the \citet{Torres:2010} sample with the parameters we infer for KELT-11.  Thus by invoking the \citet{Torres:2010} relations we are essentially extrapolating them from where they were calibrated.

Thus, given our relatively {\it a priori} unlikely inference that this is a relatively massive star (in a short phase of its lifetime), it is worth exploring what `model independent' estimates of the stellar mass we can infer.  There are two essentially direct (but not entirely independent) methods we can use to estimate the mass of the star without invoking the YY isochrones or the \citet{Torres:2010} relations.

First, we can use the estimate of the surface gravity of the star from high-resolution spectra, combined with the measurement of $R_*$ from the parallax and SED, to break the $M_*-R_*$ degeneracy.  This is problematic on its own because the surface gravities derived from spectra can have relatively large systemic errors, even with high signal-to-noise ratio spectra \citep{Torres:2012}, and furthermore spectroscopic surface gravities are also notoriously inaccurate for stars on the subgiant branch \citep{Holtzman:2015}.  Nevertheless, we can employ this process to provide a `direct' method of estimating the mass of KELT-11.  Adopting the surface gravity from the TRES spectra of \logg=$3.79 \pm 0.10$, we find $M_{*} = 1.81\pm 0.44~\msun$.  Adopting instead \logg$=3.86 \pm 0.10$ from the APF or Keck spectra, we find $M_* = 2.13 \pm 0.57~\msun$.  Thus, this procedure leads us to infer that KELT-11 is a relatively massive star.  The mass we infer is significantly higher and less precise than the global fit results, but formally does not rely on any models or empirical relations (although it does rely on the spectroscopic surface gravity).

Second, if we wish to avoid using the spectroscopic \logg\ as well as any stellar models or emprical relations, we could in principle determine the stellar mass purely through the transit light curve and the stellar radius determined from the parallax and SED.  The relation between the stellar mass and the direct observables is
\begin{equation}
\label{eqn:m*direct}
M_* = \left( \frac{4 P R_{*}^3}{\pi G T_{\rm FWMH}^3} \right) \left( \frac{\sqrt{1-e^2}}{1+e\sin{\omega}} \right)^3 (1-b^2)^{3/2},
\end{equation}
assuming $R_p \ll R_*$ and $M_p \ll M_*$.  The first factor in parentheses is reasonably well measured, and is $\sim 1.82~M_\odot$.  This depends on our estimate of $T_{\rm FWHM}$ from the global fit, but this is fairly well constrained from the follow-up photometry alone.  We are also in the process of analyzing Spitzer observations (Spitzer ID 12096) of the primary eclipse of this target, and a preliminary analysis verifies the value of $T_{\rm FWHM}$ derived in this paper.  The second factor in parentheses depends on the eccentricity, which is not well constrained.  We find that the orbit is consistent with circular, but the eccentricity could be as high as 0.16 at $1\sigma$.  Thus this factor may vary from $\sim 0.5$ to $\sim 1.3$.  Finally, the impact parameter is also poorly constrained by our data due the shallowness and duration of the transit, but is constrained to be $<0.5$ at $1\sigma$.  In fact, visual inspect of our preliminary Spitzer reduction indicates that the transit is unlikely to have an impact parameter as high as $\sim 0.5$. Nevertheless, allowing for this full range, we find that the last term can be anywhere between 1 and 0.65. Thus the full range of allowed masses is $0.6-2.4~M_{\odot}$, although we strongly favor values of $\ga 1.6~M_\odot$ based on our preliminary Spitzer data that are consistent with a nearly central transit and our strong prior that the orbit is likely to be circular.

Thus we have four methods of estimating the mass of the primary: (1) using YY isochrones, (2) using the Torres empirical relations, (3) using the spectroscopic \logg combined with an estimate of the stellar radius from the SED and parallax, and (4) using the density of the star from the transit period and duration, impact parameter, and a constraint on the eccentricity combined with the radius from the SED and parallax.  These four methods demonstrate that KELT-11 has a mass around $\ga 1.3 M_\odot$, and potentially larger.  In particular, the global modelling using the YY isochrones and the \citet{Torres:2010} relations finds good agreement, deriving a  mass of $M_{*} = 1.438^{+0.061}_{-0.052} \msun$, and $M_{*} = 1.347^{+0.072}_{-0.068} \msun$, respectively (assuming a circular orbit).  The model-independent or quasi-model-independent methods described above yield larger mass estimates but with significantly larger errors, and thus all four estimates are nominally consistent.   Although it is possible (and perhaps even likely) that there are significant systematic errors in all four of these estimates, we believe it is unlikely that those sytematics would all collude to lead us to the interpretation that this star is significantly more massive than the Sun, if, in fact, it is not.  

The fact that KELT-11 hosts a transiting planet makes it a particularly valuable member of the ``Retired A-star" sample of \citet{Johnson:2007}.  Stars with transiting planets allow an additional constraint on the properties of the host star, namely a direct measurement of $a/R_*$, and, given the (reasonable) assumption of $R_p \ll R_*$ and $M_p \ll M_*$, and a measurement or constraint on the eccentricity, an estimate of the host star mass $\rho_*$.  Currently, the constraint on the density of KELT-11 is relatively poor because of the imprecise measurement of transit duration and ingress/egress time ($T_{\rm FWHM}$ and $\tau$) due to the relatively shallow and long duration of the transit.  Nevertheless, when combined with an independent estimate
of its radius derived from the fact that it has a direct distance measurement via the Hipparcos parallax, this provides a second, purely empirical estimate of the mass of the host star.  The fact that this estimate agrees with our more model-dependent estimate of the host star mass via global fitting using isochrones, and the fact that these further agree with our estimate of the host star using the empirical relations of \citet{Torres:2010}, gives us confidence that the host star mass is indeed significantly more massive than $\sim 1.2~M_{\odot}$, and thus the host star really is a ``Retired A (or early F) star".

Once we have fully analyzed the above-mentioned Spitzer follow-up photometry of a primary transit of KELT-11b, we will have a much more precise measurement of $T_{\rm FWHM}$ and $\tau$, and thus $\rho_*$.  When combined with an exquisite measurement of the stellar parallax with Gaia \citep{Perryman:2001} and thus stellar radius, we will be able to provide much more
accurate, precise, and essentially direct empirical constraint on the host star mass, using the methods described previously.

\section{Summary and Conclusions}
\label{sec:sum}

\begin{figure}
\centering
\includegraphics[width=0.9\columnwidth,angle=0,trim={0 0 0 0},clip]{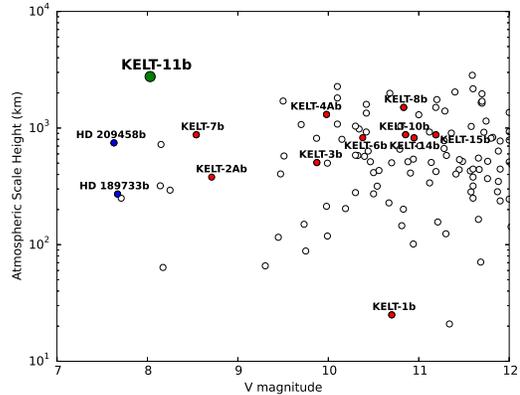}
\caption{Estimated atmospheric scale height of known transiting hot Jupiters versus the $V$-band brightness of the host star. KELT-11b is highlighted by a filled green circle, while other discoveries from the KELT survey are marked by filled red circles.  The filled blue circles mark the two well-studied benchmark hot Jupiters HD 209458b and HD 189733b. The open circles mark other known transiting exoplanets from the NASA Exoplanet Archive (accessed on 2016 May 28).}
\label{fig:ScaleHeight}
\end{figure}

KELT-11b is an extremely inflated planet (Figure \ref{fig:ScaleHeight}), with a density of just $0.093^{+0.028}_{-0.024}$ g cm$^{-3}$.  This makes KELT-11b the third lowest density planet ever discovered with a precisely measured mass and radius (those with parameter uncertainties $<$20\%).
The only comparable planets are WASP-94Ab \citep{Neveu-VanMalle:2014} and Kepler-12b \citep{Fortney:2011}, but they both orbit significantly fainter hosts.  Given its mass and level of irradiation (1.94$\times10^9$ erg/s/cm$^2$), KELT-11b has a measured radius that is about twice as large as predicted by the mass-radius-incident flux relation from \citet{Weiss:2013}. 

Another way of placing this planet in context is to note that currently there are only a handful of hot Jupiters that transit bright stars (Figure \ref{fig:ScaleHeight}). Of these KELT-11b has by far the largest atmospheric scale height, at 2763 km, assuming uniform heat redistribution and calculating the scale height along the lines of \citet{Winn:2010b}.  The ratio of scale height to planet radius is 2.8\%, with an expected size of the signal from transmission spectroscopy of 5.6\%, making KELT-11b particularly amenable to atmospheric characterization via transmission or emission spectroscopy.  The expected depth of the secondary eclipse is $1.2\alpha_{\rm therm} \times 10^{-3}$, following the calculations of \citet{Siverd:2012}.

Ultimately, the bright host star, the inflated radius, and the high equilibrium temperature make KELT-11b one of the best targets discovered to date for transmission spectroscopy.  For example, detailed studies of KELT-11b will allow its chemical composition to be determined, which in turn will constrain parameters involving its formation and evolution of planets \cite[e.g.,][]{Madhusudhan:2014}.  Furthermore, the source of inflation in hot Jupiters can be investigated in the extreme case of KELT-11b.  Future observations of KELT-11b with facilities like The Hubble Space Telescope, Spitzer, and JWST will reveal the structure and content of its atmosphere, and will set up KELT-11b as the benchmark sub-Saturn exoplanet.


It is also noteworthy that this planet has the shallowest transit depth (2.69 mmag) of any planet discovered by a ground-based transit survey, with the next-shallowest such planet having a transit depth of 3.3 mmag \citep[HAT-P-11b;][]{Bakos:2010}.  Surveys like HAT, KELT, and SuperWASP are still increasing their photometric precision, and although the TESS survey will provide higher photometric precision over the entire sky, the long time baselines of the ground-based surveys with high-quality photometry can help confirm planets with periods longer than the duration of the TESS observations.

As described in \S \ref{sec:evol}, the KELT-11 system exists in a very brief range of the host star's evolution.  The star has exhausted its core hydrogen, and is contracting such that it is about to begin (or maybe already is) undergoing shell hydrogen fusion. This stage is very short lived (roughly 60 Myr), and since transiting planets are already rare, finding one with a host star in such a stage is quite fortunate.  Once KELT-11 reaches the base of the giant branch, it will engulf KELT-11b, possibly producing a spectacular transient signal \citep{Metzger:2012}.  Thus, the detection of this one system (because it occupies such a special and short-lived period in the evolution of the star), provides an example of a direct precursor to such planet-engulfment events and transients.  It also provides an estimate of the frequency of such transient events, which can be used as a prediction for, e.g., LSST.  The recent discovery of another transiting giant planet around a subgiant star K2-39b is an additional contribution to this small sample \citep{VanEylen:2016}.
 
In addition to the potential value of KELT-11b for characterization of exoatmospheres and the frequency of planets orbiting higher-mass stars, this discovery illustrates certain aspects of the current state of transit discovery.  This planet was discovered due to the combination of both transit and RV survey data.  The KELT survey observations (\S \ref{sec:ks}) and follow-up photometry (\S \ref{sec:Follow-up_Photometry}) enabled us to identify this target as a good candidate, but with such a low mass planet, our typical follow-up methods to obtain an RV orbit would have been extremely hard-pressed to enable dynamical confirmation purely through follow-up RV observations.  However, the addition of the CPS survey data provided the evidence that this was a real planet, prompting us to gather the additional APF observations to enable reliable confirmation.  Furthermore, the CPS RV observations by themselves were not sufficient to verify HD 93396 as a planet host without the accompanying transit evidence from KELT.  We believe that this synergy between multiple types of survey data will be of great value over the next several years, especially with the expected launch of the TESS mission and availability of nearly all-sky high precision photometry.

\acknowledgements

Work by B.S.G. and D.J.S. was partially supported by NSF CAREER Grant AST-1056524.  M.B. is supported by a NASA Space Technology Research Fellowship.  J.P. would like to thank Stephen Faulkner and Katheryn Kirkwood for help during a difficult time.  D.W.L. acknowledges partial support from the Kepler Extended Mission under Cooperative Agreement NNX13AB58A with the Smithsonian Astrophysical Observatory.  D. B. acknowledges financial support from the National Centre for Competence in Research PlanetS supported by the Swiss National Science Foundation.  T.B. was partially supported by funding from the Center for Exoplanets and Habitable Worlds. The Center for Exoplanets and Habitable Worlds is supported by the Pennsylvania State University, the Eberly College of Science, and the Pennsylvania Space Grant Consortium.

B.J.F. notes that this material is based upon work supported by the National Science Foundation Graduate Research Fellowship under grant No. 2014184874. Any opinion, findings, and conclusions or recommendations expressed in this material are those of the authors(s) and do not necessarily reflect the views of the National Science Foundation.

This work has made use of NASA’s Astrophysics Data System, the Extrasolar Planet Encyclopedia at exoplanet.eu \citep{Schneider:2011}, the SIMBAD database operated at CDS, Strasbourg, France, and the VizieR catalogue access tool, CDS, Strasbourg, France \citep{Ochsenbein:2000}. 

Certain calculations in this paper were carried out on the Ruby cluster operated by the Ohio Supercomputer Center \citep{Center:1987}.

MINERVA is a collaboration among the Harvard-Smithsonian Center for Astrophysics, The Pennsylvania State University, University of Montana, and University of New South Wales. MINERVA is made possible by generous contributions from its collaborating institutions and Mt. Cuba Astronomical Foundation, The David \& Lucile Packard Foundation, National Aeronautics and Space Administration (EPSCOR grant NNX13AM97A), and The Australian Research Council (LIEF grant LE140100050).

This publication makes use of data products from the Widefield Infrared Survey Explorer, which is a joint project of the University of California, Los Angeles, and the Jet Propulsion Laboratory/California Institute of Technology, funded by the National Aeronautics and Space Administration.

This publication makes use of data products from the Two Micron All Sky Survey, which is a joint project of the University of Massachusetts and the Infrared Processing and Analysis Center/California Institute of Technology, funded by the National Aeronautics and Space Administration and the National Science Foundation.

This paper makes use of data from the first public release of the WASP data \citep{Butters:2010} as provided by the WASP consortium and services at the NASA Exoplanet Archive \citep{Akeson:2013}, which is operated by the California Institute of Technology, under contract with the National Aeronautics and Space Administration under the Exoplanet Exploration Program, the Exoplanet Orbit Database and the
Exoplanet Data Explorer at exoplanets.org \citep{Han:2014}.

\bibliographystyle{apj}

\bibliography{KELT-11b}

\end{document}